\documentclass[sigconf]{acmart}
\AtBeginDocument{%
  }

\copyrightyear{2026}
\acmYear{2026}
\setcopyright{cc}
\setcctype{by}
\acmConference[KDD '26]{Proceedings of the 32nd ACM SIGKDD Conference on Knowledge Discovery and Data Mining V.2}{August 09--13, 2026}{Jeju Island, Republic of Korea}
\acmBooktitle{Proceedings of the 32nd ACM SIGKDD Conference on Knowledge Discovery and Data Mining V.2 (KDD '26), August 09--13, 2026, Jeju Island, Republic of Korea}
\acmDOI{10.1145/3770855.3817578}
\acmISBN{979-8-4007-2259-2/2026/08}

\usepackage{multirow}
\usepackage{rotating}   
\usepackage{makecell}   
\usepackage{xspace} 
\usepackage{enumitem} 

\newcommand{\bench}{FinStressTS}
\begin{document}

\title{FinStressTS: A Parametric Synthetic Benchmark for Time-Series Forecasting in Finance}

\author{Jiaze Sun}
\affiliation{%
  \institution{National University of Singapore}
  \country{Singapore}
}
\email{e0564914@u.nus.edu}

\author{Kelvin J.L. Koa}
\authornote{Corresponding author.}
\affiliation{%
  \institution{National University of Singapore}
  \department{Asian Institute of Digital Finance}
  \country{Singapore}
}
\email{kelvin.koa@u.nus.edu}

\author{Ruiyang Ni}
\affiliation{%
  \institution{Nanyang Technological University}
  \country{Singapore}
}
\email{rni001@e.ntu.edu.sg}

\author{Yize Liu}
\affiliation{%
  \institution{National University of Singapore}
  \country{Singapore}
}
\email{liuyize51@u.nus.edu}

\author{Haonan Chen}
\affiliation{%
  \institution{National University of Singapore}
  \country{Singapore}
}
\email{e1538433@u.nus.edu}

\author{Ke-Wei Huang}
\affiliation{%
  \institution{National University of Singapore}
  \department{Asian Institute of Digital Finance}
  \country{Singapore}
}
\email{dishkw@nus.edu.sg}

\renewcommand{\shortauthors}{Jiaze Sun et al.}


\begin{abstract}
Financial forecasting is notoriously difficult due to low signal-to-noise ratios, latent factors, and discontinuous jumps. A critical limitation of real-world benchmarks is the inability to attribute failure: researchers see that a model underperforms, but cannot isolate why, as the generative mechanisms are unobservable entanglements of noise and structure. Furthermore, real financial data reveal only one realized path, making it difficult to assess tail-risk calibration or data efficiency under controlled conditions. We introduce FinStressTS, a mechanism-aware synthetic benchmark designed to bridge this gap by providing controlled environments that link model behavior directly to specific structural causes.

FinStressTS comprises 30 diagnostic environments organized around six canonical mechanism families: volatility clustering, multi-scale persistence, heavy-tailed shocks, regime switching, self-exciting jumps, and zero-inflated processes. We evaluate models on two distinct tasks: (1) Point Forecasting, where we assess predictive accuracy using NMAE across five diagnostic settings; and (2) Probabilistic Forecasting, where we benchmark distributional calibration using CRPS in settings with known data-generating mechanisms—an evaluation difficult to isolate with real data. We rigorously evaluate 15 diverse models, ranging from classical econometric methods (HAR, VAR) to modern Transformer-based forecasters (PatchTST, iTransformer) and deep probabilistic architectures (DeepAR, TSFlow), while also conducting learning curve analyses to measure performance as a function of sample size.

Our evaluation reveals critical insights: (1) Mechanism Sensitivity: Model performance is governed by architectural inductive bias rather than capacity—simple autoregressive and linear models consistently outperform Transformers in volatility-, tail-, and jump-driven environments;
(2) Distributional Alignment: Parametric probabilistic models (e.g., DeepAR) achieve superior calibration in stationary settings by aligning with volatility dynamics, whereas flexible density models (e.g., flows and mixtures) are necessary when predictive distributions become multimodal or sparse;
(3) Data Inefficiency: Neural models require substantially more data to match simple baselines and often fail to benefit from additional samples in stationary regimes, with larger datasets helping primarily when learning latent regime structure or complex predictive distributions. FinStressTS provides the community with an open framework to diagnose failure modes and advance risk-aware forecasting; the accompanying code is available on GitHub.\footnotemark

\end{abstract}




\maketitle
\footnotetext{Code available at: \url{https://github.com/jiazeee/FinStressTS}.}

\section{Introduction}
\label{sec:intro}

Financial time-series forecasting is a cornerstone of modern data science, yet it remains notoriously difficult due to extremely low signal-to-noise ratios and complex, evolving dynamics. Recent large-scale evaluations, such as the M6 Financial Forecasting Competition \cite{makridakis2024m6}, have demonstrated that sophisticated deep learning models frequently fail to outperform simple heuristics when evaluated on raw market returns. While this highlights the inherent unpredictability of financial markets, it also exposes a fundamental methodological paradox: real-world benchmarks offer authenticity but function as "black boxes" for diagnosis. Financial data entangles multiple latent mechanisms simultaneously—such as volatility clustering, heavy-tailed shocks, regime shifts, and self-exciting jumps—making it impossible to isolate the specific structural reasons for a model's underperformance. When a forecaster fails, researchers cannot definitively determine whether the error stems from distributional miscalibration, an inability to adapt to regime changes, or insufficient data efficiency. This attribution gap limits what current benchmarks can verify: we can measure that modern architectures fail, but we cannot rigorously determine why.

Current evaluation infrastructure reinforces this gap by forcing researchers to choose between authenticity and diagnostic control. Reliance on historical data restricts evaluation to a single, unrepeatable realization of a stochastic process, which precludes the systematic assessment of robustness and scalability that is essential for deployment. For instance, with real data, one cannot generate "counterfactual" histories to test how a model’s performance degrades under increasingly "dirty" conditions (e.g., noisy inputs, outliers) or rigorously measure data efficiency (learning curves) without the confounding effects of time-varying regimes. Meanwhile, alternative benchmarks fail to bridge this divide: generic time-series datasets lack the distinctive risk profile of financial markets, while existing synthetic baselines frequently rely on oversimplified Gaussian assumptions that fail to challenge modern architectures. Consequently, the community lacks a "middle ground" that combines parametric control with financial realism to enable the "green-box" testing impossible with real data.

To provide this missing diagnostic ``middle ground,'' we introduce FinStressTS, a mechanism-aware synthetic benchmark for financial time-series forecasting. Rather than attempting to reproduce the full complexity of markets, FinStressTS follows a controlled stress-testing philosophy: it isolates six canonical mechanism families that correspond to widely documented stylized facts of asset returns \cite{cont2001stylized}, implements each family via parametrically controlled data-generating processes grounded in econometric theory, and systematically varies their structural parameters to construct 30 diagnostic environments with controlled mechanism-specific settings. Because the generative mechanisms are known by design, FinStressTS enables evaluations that are infeasible with historical benchmarks: (i) \textit{probabilistic calibration} can be assessed using proper scoring rules such as CRPS in settings where the data-generating mechanism is known \cite{gneiting2007scoring}, and (ii) \textit{data efficiency} can be measured through learning curves that quantify how much data different model classes require to reliably capture specific financial dynamics. In this way, FinStressTS turns opaque underperformance into interpretable diagnosis—linking errors in accuracy, robustness, and calibration to precise structural causes rather than confounded explanations.

FinStressTS is organized around six mechanism families that distinguish financial time series from generic forecasting domains and jointly capture core empirical regularities. \textit{(i) Volatility clustering} models persistent conditional heteroskedasticity through ARCH/GARCH dynamics \cite{engle1982arch,bollerslev1986garch}. \textit{(ii) Multi-scale volatility persistence} captures long-memory effects across daily/weekly/monthly horizons via the HAR structure \cite{corsi2009har}. \textit{(iii) Heavy-tailed shocks and outliers} reproduce fat-tailed return behavior \cite{cont2001stylized} using Student-$t$ innovations (e.g., in conditional volatility settings) \cite{bollerslev1987garch}. \textit{(iv) Regime switching} represents abrupt, persistent shifts in market conditions using Markov-switching processes \cite{hamilton1989new}. \textit{(v) Self-exciting jump dynamics} capture event clustering—where extreme events raise the likelihood of subsequent jumps—via Hawkes-type processes \cite{hawkes1971hawkes}. \textit{(vi) Zero-inflated processes} represent intermittent activity with extended zero/near-zero periods punctuated by bursts, which is common in illiquid assets and transaction-level settings \cite{lambert1992zip}. For each family, FinStressTS defines multiple diagnostic levels by varying mechanism-specific parameters (e.g., persistence, tail heaviness, regime duration, jump intensity, zero inflation), enabling targeted failure-mode analysis under isolated mechanisms as well as realistic compound stress scenarios.

We validate FinStressTS with a comprehensive benchmarking study of 15 forecasting models spanning classical statistical baselines (e.g., AR/HAR/VAR), representative deep point forecasters (including Transformer variants such as PatchTST and iTransformer), and deep probabilistic architectures (e.g., autoregressive likelihood models and flow-based generative forecasters). We evaluate both point accuracy and distributional quality using proper scoring rules in controlled settings with known data-generating mechanisms \cite{gneiting2007scoring}, and we further conduct learning-curve analyses to quantify data efficiency under controlled mechanism stresses. This setup yields insights that black-box historical benchmarks cannot isolate. First, \textit{limited mean predictability}: across many environments, simple autoregressive-style baselines are competitive—and often superior—to complex neural models, emphasizing that increasing volatility complexity does not automatically translate into learnable improvements in expected returns. Second, \textit{mechanism sensitivity}: architectures that perform well under stationary volatility clustering can degrade sharply under regime-switching or jump-driven environments, revealing failure modes that are obscured when mechanisms are entangled. Third, \textit{probabilistic fragility}: deep probabilistic models can exhibit systematic miscalibration under heavy tails and stress dynamics, whereas models with more flexible distributional families can better preserve calibration. Finally, \textit{data inefficiency}: learning curves reveal that neural methods often require multiples more data than classical baselines to achieve stable performance and reliable calibration. Overall, FinStressTS acts as a diagnostic instrument rather than a leaderboard: it turns aggregate errors into mechanism-level failure signatures.

This paper makes the following contributions:
\begin{enumerate}[label=(\arabic*), leftmargin=*, nosep]
    \item A mechanism-aware benchmark comprising 30 diagnostic environments with parametric control over six canonical financial mechanisms, each grounded in established econometric theory and empirical stylized facts.
    \item Evaluation of 15 forecasting models across point and probabilistic tasks, distinguishing between verifiable point accuracy and distributional calibration: unlike historical data where true distributions and volatility dynamics are unobservable, our benchmark enables evaluation under known data-generating mechanisms.
    \item Systematic learning curve analyses that quantify data efficiency requirements across mechanism families, revealing how sample size impacts model reliability in ways obscured by standard train-test splits.
    \item Open-source implementation with reproducible experimental protocols, enabling the community to conduct targeted stress tests, diagnose failure modes, and validate architectural claims under controlled conditions.
    \item Empirical insights on mechanism-specific vulnerabilities that inform principled model selection for high-stakes financial applications.
\end{enumerate}
\section{Related Work}
\label{sec:related}

\subsection{Point and Probabilistic Time-Series Forecasting}
\label{sec:related_forecasting}

Research on time-series forecasting spans classical statistical models and modern deep learning architectures, with growing emphasis on both point prediction and full predictive distributions. We organize the discussion around three strands that align with the models evaluated in this work: (i) classical point forecasters, (ii) neural point forecasters, and (iii) probabilistic time-series models.

\paragraph{Classical point forecasting.}
Classical econometric models remain strong baselines in financial forecasting due to their interpretability and robustness in low-data regimes. Linear autoregressive (AR) models and vector autoregressions (VAR) capture temporal and cross-variable dependencies through lagged relationships and are widely used in forecasting and policy analysis \cite{box1976analysis,lutkepohl2013introduction,cont2001empirical}. To capture multi-scale dynamics in financial volatility, the Heterogeneous Autoregressive model of realized volatility (HAR-RV) regresses future realized volatility on daily, weekly, and monthly realized-volatility components, providing a parsimonious approximation to long-memory effects \cite{corsi2009simple,andersen2003modeling}. Despite their structural simplicity, these models often remain competitive and provide useful reference points for benchmarking \cite{makridakis2018m4}.

\paragraph{Neural point forecasting.}
Recent work has developed a diverse set of neural architectures for time-series forecasting. A first family emphasizes simplicity and decomposition, exemplified by DLinear, which decomposes an input series into trend and residual seasonal components before applying linear mappings \cite{zeng2023transformers}. A second family adapts transformers to time series via specialized inductive biases, including Fourier enhanced structure (FEDformer) \cite{zhou2022fedformer} and series decomposition with auto-correlation (Autoformer) \cite{wu2021autoformer}. A third family focuses on efficiency and structure-aware tokenization, such as PatchTST, which operates on local temporal patches \cite{nie2022time}, and iTransformer, which inverts the standard tokenization by embedding each variate's historical sequence as a token and applying attention across variates rather than temporal tokens \cite{liu2023itransformer}. More recent models, including TimeXer \cite{wang2024timexer}, explicitly model cross-time and cross-variate feature interactions, and the Nonstationary Transformer \cite{liu2022non} introduces learned de-stationarization to better handle distributional shifts. Collectively, these models provide representative and competitive neural point-forecasting baselines.

\paragraph{Probabilistic time-series forecasting.}
Beyond point prediction, probabilistic forecasting aims to model full predictive distributions, capturing uncertainty, tail risk, and dependence structure \cite{gneiting2014probabilistic}. 
DeepAR is a seminal deep autoregressive model that parameterizes a likelihood conditioned on past observations and covariates, enabling scalable probabilistic forecasting across multiple related series \cite{salinas2020deepar,alexandrov2020gluonts}. 
Subsequent work has developed more flexible generative and distributional models for multivariate forecasting. 
TimeGrad introduces autoregressive denoising diffusion models for multivariate probabilistic time-series forecasting, sampling future values through a learned reverse diffusion process \cite{rasul2021autoregressive}. 
TSFlow uses conditional flow matching with Gaussian-process priors and optimal-transport paths to model time-series distributions and enable probabilistic forecasting \cite{kollovieh2024flow}. 
TimeMCL adapts Multiple Choice Learning to multivariate forecasting, using multiple prediction heads and a Winner-Takes-All loss to generate diverse plausible futures \cite{cortes2025winner}. 
More recently, RATD augments diffusion-based forecasting with an embedding-based retrieval module, using retrieved reference series to guide the denoising process \cite{liu2024retrieval}, while QuantileFormer formulates Transformer-based probabilistic forecasting through pattern-mixture decomposition, combining quantile drift, divergence patterns, and Gaussian mixture components \cite{shao2025quantileformer}. 
Evaluation typically relies on proper scoring rules such as the Continuous Ranked Probability Score (CRPS) \cite{gneiting2014probabilistic}.

\paragraph{Relation to our study.}
We benchmark classical econometric baselines (AR, HAR, VAR), modern neural point forecasters (DLinear, Autoformer, FEDformer, PatchTST, iTransformer, TimeXer, and Nonstationary Transformer), and representative probabilistic models (DeepAR, TimeGrad, TSFlow, TimeMCL, RATD, and QuantileFormer). This enables systematic comparison across traditional versus neural methods and point versus probabilistic paradigms under controlled financial mechanisms within a unified evaluation framework.

\subsection{Time-Series Benchmarking and Evaluation}
\label{sec:related_benchmark}

Benchmarking has played a central role in advancing time-series forecasting (TSF) by providing standardized datasets, evaluation protocols, and reproducible baselines \cite{makridakis2018m4,godahewa2021monash}. 
A wide range of general-purpose tools and libraries have been developed for forecasting research and practice, including Prophet for decomposable business forecasting \cite{taylor2018forecasting}, sktime for unified time-series learning \cite{loning2019sktime}, TSlib for neural forecasting models \cite{wang2024tssurvey}, and GluonTS and NeuralForecast for probabilistic forecasting and evaluation \cite{alexandrov2020gluonts,olivares2022library_neuralforecast}. More recently, ProbTS has sought to unify point and probabilistic evaluation and integrate emerging time-series foundation models \cite{zhang2024probts}. 
These toolkits have significantly improved reproducibility and accessibility in TSF research. However, they are largely domain-agnostic and are not designed around financial-market stress mechanisms.

In terms of datasets and benchmarks, early collections such as M3 and M4 focused primarily on univariate forecasting across heterogeneous domains \cite{makridakis2000m3,makridakis2018m4}, while the Monash Time Series Forecasting Archive further consolidated diverse public datasets for evaluating global forecasting algorithms \cite{godahewa2021monash}. 
More recent benchmarks and libraries have increasingly emphasized multivariate and long-horizon forecasting, including the long-term forecasting benchmarks popularized by LTSF-Linear and unified multivariate evaluation pipelines such as BasicTS+ \cite{zeng2023transformers,shao2024exploring}. 
As highlighted by TFB \cite{qiu2024tfb}, existing TSF benchmarks exhibit three recurring issues: insufficient coverage of data domains, stereotype bias against traditional statistical methods, and inconsistent or inflexible evaluation pipelines. 
TFB addresses these concerns by expanding domain coverage, including traditional baselines, and standardizing evaluation protocols. 
Nevertheless, such benchmarks are primarily designed for broad empirical comparison across heterogeneous domains. 
Financial time series, when included, typically constitute one domain among many rather than the organizing focus of the benchmark.

A complementary line of work has explored finance-oriented datasets aimed at improving the evaluation of time-series models on financial tasks. Prior studies in financial time series modeling often assess models on different sliced historical data from specific markets \cite{wang2022adaptive, chen2024automatic, duan2022factorvae, zeng2024trade}. To address the lack of consistency across such evaluations, FinTSB \cite{hu2025fintsb} constructed consolidated datasets using 15 years of historical stock data. While this approach improves consistency in time horizons and market coverage, it still relies on real financial data that must be continuously updated. Moreover, real financial data entangle multiple stochastic mechanisms simultaneously, making it difficult to attribute model performance to specific structural features or to conduct controlled stress tests.

To obtain greater experimental control, several prior works have explored synthetic or semi-synthetic datasets. Examples include simulation-based generators such as GRATIS \cite{kang2020gratis}, synthetic benchmark suites for generative time-series modeling such as TSGM \cite{nikitin2024tsgm} and TSGBench \cite{ang2023tsgbench}, and task-specific synthetic datasets such as TimeGraph \cite{ferdous2025timegraph} for causal discovery. In parallel, recent work has applied generative models to produce synthetic financial time series, including GAN-based approaches such as TimeGAN \cite{ozturk2024enhancing} and QuantGAN \cite{wiese2020quant}. These methods focus on data realism, augmentation, or generative modeling, but are not designed as mechanism-controlled benchmarks for forecasting evaluation.

Therefore, our work introduces a mechanism-aware, parametric synthetic benchmark for financial time-series forecasting. By organizing six interpretable mechanism families with controllable diagnostic levels, the benchmark enables targeted stress testing, failure-mode analysis, and data-efficiency studies across models within a unified evaluation framework.

\subsection{Financial Econometrics and Mechanism Modeling}
\label{sec:related_econ}

Financial econometrics provides the theoretical foundation for our mechanism-aware design by documenting the structural properties that fundamentally shape forecasting difficulty. Canonical stylized facts include \emph{volatility clustering} and \emph{multi-scale persistence} \cite{engle1982autoregressive,corsi2009har}, which create time-varying risk; \emph{heavy tails} and \emph{regime shifts} \cite{cont2001empirical,hamilton1989new}, which violate Gaussian stationarity; and \emph{jump dynamics}—both self-exciting \cite{hawkes1971hawkes} and zero-inflated \cite{andersen2007roughing}—which introduce discontinuity and event clustering. 
Finally, markets exhibit strong \emph{cross-sectional dependence} governed by latent factor structures \cite{fama1993common,ross2013arbitrage}. 
\bench{} builds directly on these insights by operationalizing these mechanisms into parametric data-generating processes, enabling controlled evaluation under financially realistic yet diagnostically transparent conditions.
\section{The \bench\ Stress-Test Suite}
\label{sec:suite}

\subsection{Design principles}
\label{sec:design_principles}
We design \bench\ as a mechanism-aware stress-test suite for financial time-series forecasting, guided by four core principles:

\begin{enumerate}[label=(\arabic*), leftmargin=*, nosep]
  \item \textbf{Econometric fidelity.} 
  All six mechanisms are grounded in canonical stochastic processes that are well-established in theory and empirically validated across decades of econometric literature. Unlike arbitrary synthetic data, each case targets a specific, universally recognized stylized fact of asset returns, ensuring that the benchmark rigorously reflects the theoretical and empirical consensus on how financial markets evolve.

  \item \textbf{Diagnostic control.} 
  Our parametric approach allows us to explicitly control the strength of specific time-series properties—such as the magnitude of jumps or the frequency of regime switches.
  This enables targeted stress testing: by varying these diagnostic control parameters, we can rigorously evaluate how well data-driven algorithms capture specific structural patterns and identify the exact threshold at which they fail.

  \item \textbf{Verifiable ground truth.}
  Because the data-generating process is fully specified, \bench\ provides access to the true conditional distribution $\mathbb{P}(y_{t+1}|\mathcal{H}_t)$. 
  This enables a rigorous evaluation of probabilistic calibration and tail risk—assessments that are theoretically impossible with real-world data where the true distribution is unknown.

  \item \textbf{Multivariate panel structure.}  
  Each environment generates a panel of return series whose co-movement is driven by shared latent drivers, such as common factors, regimes, or market-wide jumps, rather than by an arbitrary covariance matrix. This makes cross-sectional dependence interpretable and controllable: by varying the strength of common factors, heterogeneity in factor loadings, and idiosyncratic noise, \bench\ tests whether multivariate models can exploit shared structure across series.
\end{enumerate}

Together, these principles ensure that \bench\ supports controlled, interpretable, and reproducible evaluation, enabling the community to move beyond leaderboard rankings toward a diagnostic understanding of the forecasting robustness.

\subsection{Mechanism families}
\label{sec:six_cases}
\begin{figure*}[t]
  \centering
  \includegraphics[width=\textwidth]{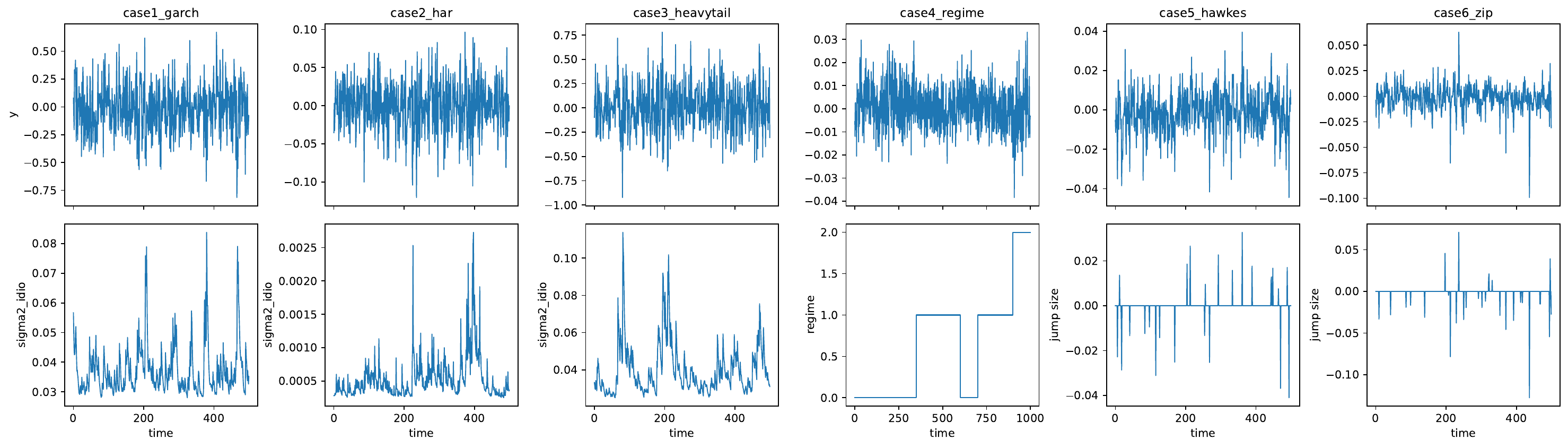}
  \caption{Illustrative examples of synthetic time series generated by each mechanism family under the Level~1 diagnostic configuration. Each column corresponds to one mechanism. The top row displays the observed series, and the bottom row displays the associated latent process governing conditional variance, regime state, or jump magnitude.}
  \Description{A 2-by-6 grid of time-series plots. Each column corresponds to one synthetic data mechanism. The top row shows the generated time series, and the bottom row shows the associated latent process, such as volatility, regime, or jump size.}
  \label{fig:examples}
\end{figure*}

We instantiate six mechanism families and each family isolates a distinct stochastic mechanism, supporting multivariate panel generation and interpretable diagnostic control (Figure~\ref{fig:examples}).

\begin{enumerate}[label=(\arabic*), leftmargin=*, nosep]
\item \textbf{Volatility clustering (GARCH-type).}
\textbf{Finance motivation:} conditional heteroskedasticity and volatility persistence are the most robust empirical regularities in returns \cite{engle1982arch,bollerslev1986garch}. 
\textbf{Diagnostic question:} can a model adapt its risk over time and avoid systematically under/overestimating volatility during clusters of large moves?
\textbf{Controls:} persistence strength and mean reversion, cross-sectional heterogeneity, and the signal-to-noise ratio between systematic (factor) and idiosyncratic risk.

\item \textbf{Multi-scale volatility persistence (HAR-type).}
\textbf{Finance motivation:} realized volatility exhibits persistence across multiple horizons and long-memory-like behavior \cite{andersen2003rv,corsi2009har}.
\textbf{Diagnostic question:} can a model learn \emph{multi-scale} temporal dependence (daily/weekly/monthly components) rather than overfitting short-term fluctuations?
\textbf{Controls:} overall persistence, the relative weight of long-horizon components (memory decay), and baseline noise levels.

\item \textbf{Heavy tails and outliers.}
\textbf{Finance motivation:} return distributions are fat-tailed and contain extreme observations far more frequently than Gaussian models predict \cite{cont2001stylized,embrechts1997extremes}. 
\textbf{Diagnostic question:} do forecasts remain reliable under tail risk, and are predictive intervals/calibration robust when shocks are heavy-tailed or contaminated?
\textbf{Controls:} tail heaviness (degrees of freedom), contamination frequency, and outlier magnitude.

\item \textbf{Regime switching with structural breaks.}
\textbf{Finance motivation:} markets undergo persistent shifts (e.g., crises) that violate global stationarity assumptions. 
\textbf{Diagnostic question:} can a model detect and adapt to distribution shift without catastrophic degradation?
\textbf{Controls:} regime persistence and break frequency, regime separation (identifiability), and within-regime temporal dependence.

\item \textbf{Self-exciting jumps (Hawkes-type).}
\textbf{Finance motivation:} news shocks often cluster in time, producing cascades rather than independent jumps; Hawkes processes are a standard model for such self-excitation in finance \cite{hawkes1971hawkes,bacry2015hawkes}. 
\textbf{Diagnostic question:} can a model remain calibrated and responsive when shocks arrive in bursts and stress propagates across assets?
\textbf{Controls:} excitation strength (criticality), memory decay, cross-sectional coupling via shared jump components, and jump magnitude distributions.

\item \textbf{Zero-inflated jumps (sparse activity).}
\textbf{Finance motivation:} illiquid assets and transaction-level series exhibit long inactive stretches punctuated by bursts of activity, producing zero-inflation and intermittent dynamics \cite{lambert1992zip,lesmond1999costs}. 
\textbf{Diagnostic question:} can a model handle sparse-event regimes without collapsing calibration?
\textbf{Controls:} sparsity rate, conditional arrival intensity in the active state, and jump magnitude distributions.
\end{enumerate}

A summary of these families and control parameters is provided in Table~\ref{tab:case_summary}.
Full specifications are deferred to Appendix~\ref{sec:appendix_dgp}.

\begin{table}[t]
\centering
\caption{Summary of \bench\ mechanism families. Each case targets a specific financial stylized fact and exposes interpretable parameters for diagnostic control.}
\label{tab:case_summary}
\small
\setlength{\tabcolsep}{4pt}
\begin{tabular}{c p{2.6cm} p{4.2cm}}
\toprule
\textbf{Case} & \textbf{Mechanism} & \textbf{Diagnostic Parameters (Control Knobs)} \\
\midrule
1 & Volatility clustering 
  & Persistence ($\rho$); Heterogeneity; SNR \\
2 & Multi-scale persistence 
  & Long-memory share \\
3 & Heavy tails / Outliers 
  & Tail d.o.f. ($\nu$); Outlier prob.; Magnitude \\
4 & Regime switching 
  & Break frequency; Regime separation; Memory \\
5 & Self-exciting jumps 
  & Excitation ($\alpha$); Decay ($\beta$); Criticality \\
6 & Zero-inflated jumps 
  & Sparsity ($\pi$); Active intensity; Magnitude \\
\bottomrule
\end{tabular}
\end{table}

\subsection{Diagnostic level design}
\label{sec:levels} 
For each mechanism family, we define five diagnostic levels that systematically vary structural parameters to enable targeted failure-mode analysis. Rather than imposing monotonic difficulty, each level isolates a specific mechanism dimension \emph{ceteris paribus}: Level 1 provides a balanced baseline for standardized comparison and data-efficiency experiments, while Levels 2--5 each vary specific parameters to stress-test distinct structural properties (e.g., regime duration vs. regime separation vs. within-regime persistence). This design enables direct attribution of performance changes to precise causal factors, transforming opaque failures into interpretable diagnoses. Complete parameter specifications for all 30 environments are provided in the accompanying repository.
\section{Models and Evaluation Protocol}
\label{sec:models_protocol}

We evaluate 15 models along five structural dimensions.
\begin{enumerate}[label=(\arabic*), leftmargin=*, nosep]
    \item \textbf{Linear vs. Nonlinear:} 
    We compare linear and low-capacity baselines (AR, HAR, VAR, DLinear) with deep nonlinear architectures.

    \item \textbf{Marginal vs. Cross-Series Information Use:} 
    We contrast models that explicitly mix cross-series information (VAR, iTransformer, TimeXer) with channel-independent or marginal forecasting models (e.g., PatchTST, DeepAR).

    \item \textbf{Stationarity Handling:} 
    We compare models with explicit decomposition, normalization, or stationarization mechanisms (DLinear, Autoformer, FEDformer, Non-stationary Transformer) against other attention-based forecasters to test robustness to trends and regime shifts.

    \item \textbf{Probabilistic Family:} 
    We compare likelihood-based forecasting (DeepAR), flow-matching and diffusion-style generative forecasting (TSFlow, RATD), multiple-choice sample generation (TimeMCL), and quantile-structured probabilistic modeling (QuantileFormer).

    \item \textbf{Dependence Modeling:} 
    We distinguish marginal probabilistic forecasts from joint multivariate forecast samples for 
    $P(\mathbf{y}_{t+1}\mid\mathcal{H}_t)$, enabling evaluation of aggregate risk calibration.
\end{enumerate}

\subsection{Forecasting Task and Protocol}
\label{sec:task_definition}

\paragraph{\noindent \textbf{Data Generation and Splitting.}} For each mechanism, we generate a panel of $N=50$ series and $T_{\text{total}}=2,000$ steps. We use a strict chronological split (60\% training, 20\% validation, 20\% testing). Standardization ($z$-score) is applied per series $i$ using statistics computed solely on the training split.

\paragraph{\noindent \textbf{Rolling One-Step-Ahead Forecasting.}} We adopt a rolling setting ($H=1$). At time $t$, the model observes a lookback window $\mathbf{Y}_{t-L+1:t} \in \mathbb{R}^{N \times L}$ (where $L=96$) containing histories for all series. Parameters are fixed after training; no online updates are performed.

\paragraph{\noindent \textbf{Point Forecasting Task.}} A point forecaster learns a mapping $f_\theta: \mathbb{R}^{N \times L} \to \mathbb{R}^N$ to predict the vector of next-step values:
\begin{equation}
\hat{\mathbf{y}}_{t+1}
=
[\hat{y}_{1,t+1}, \dots, \hat{y}_{N,t+1}]^\top
=
f_\theta\left(\mathbf{Y}_{t-L+1:t}\right).
\end{equation}
Models capable of global modeling process the full panel jointly; univariate models process each series $i$ independently to output $\hat{y}_{i,t+1}$.

\paragraph{\noindent \textbf{Probabilistic Forecasting Task.}} Probabilistic models estimate the conditional distribution of the next step given history $\mathcal{H}_t$:
\begin{equation}
\hat{F}_{t+1}(\mathbf{z} \mid \mathcal{H}_t)
\approx
\mathbb{P}(Y_{1,t+1} \le z_1,\dots,Y_{N,t+1} \le z_N \mid \mathcal{H}_t).
\end{equation}

Depending on the architecture, this predictive distribution may factorize across series as
$\prod_{i=1}^N \hat{F}_{i,t+1}(z_i \mid \mathcal{H}_t)$
or model cross-series dependencies explicitly. Evaluation uses $S=100$ Monte Carlo samples $\{\hat{\mathbf{y}}_{t+1}^{(s)}\}_{s=1}^S$ drawn from $\hat{F}_{t+1}$.

\begin{table*}[t]
\centering
\tiny
\setlength{\tabcolsep}{2.4pt}
\renewcommand{\arraystretch}{0.90}

\caption{Point forecasting results (NMAE $\downarrow$). Rows are (case, level). Best is \textbf{bold}; second best is \underline{underlined}.}
\label{tab:main_results_point}

\scalebox{1.4}{%
\begin{tabular}{llp{2.1cm}*{11}{p{0.68cm}}}
\toprule
\multirow{2}{*}{\textbf{Case}} &
\multirow{2}{*}{\textbf{Level}} &
\multirow{2}{*}{\textbf{Scenario}} &
\multicolumn{11}{c}{\textbf{Point models (NMAE $\downarrow$)}} \\
\cmidrule(lr){4-14}
& & &
\textbf{Naive} &
\textbf{AR(1)} &
\textbf{HAR} &
\textbf{VAR} &
\textbf{DLinear} &
\textbf{PatchTST} &
\textbf{iTrans} &
\textbf{AutoF} &
\textbf{FEDF} &
\textbf{NTrans} &
\textbf{TimeXer} \\
\midrule

\multirow{5}{*}{\textbf{1}} & L1 & Baseline
& 1.1267 & \textbf{0.7970} & \underline{0.7989} & 0.8105 & 0.7980 & 0.8065 & 0.8096 & 0.8629 & 0.8472 & 0.8093 & 0.8040 \\
& L2 & Factor persistence
& 1.1054 & \underline{0.8002} & 0.8009 & 0.8178 & \textbf{0.7971} & 0.8069 & 0.8150 & 0.8519 & 0.8453 & 0.8070 & 0.8082 \\
& L3 & Idio persistence
& 1.1122 & \textbf{0.7870} & \underline{0.7881} & 0.8037 & 0.7883 & 0.7945 & 0.7994 & 0.8527 & 0.8439 & 0.7965 & 0.7949 \\
& L4 & Heterogeneity
& 1.1137 & \textbf{0.7972} & \underline{0.7976} & 0.8147 & 0.7993 & 0.8122 & 0.8232 & 0.8593 & 0.8507 & 0.8077 & 0.8129 \\
& L5 & Low SNR
& 1.1197 & \underline{0.7942} & 0.7947 & 0.8124 & \textbf{0.7940} & 0.8002 & 0.8012 & 0.8666 & 0.8517 & 0.8037 & 0.7996 \\
\midrule

\multirow{5}{*}{\textbf{2}} & L1 & Baseline
& 1.1277 & \textbf{0.7881} & \underline{0.7886} & 0.8088 & 0.7903 & 0.8042 & 0.8151 & 0.8507 & 0.8449 & 0.7981 & 0.8062 \\
& L2 & High persistence
& 1.0170 & \textbf{0.7386} & \underline{0.7393} & 0.7620 & 0.7394 & 0.7508 & 0.7540 & 0.8152 & 0.7962 & 0.7488 & 0.7464 \\
& L3 & Long memory
& 1.1126 & \textbf{0.7900} & \underline{0.7909} & 0.8132 & 0.7917 & 0.8033 & 0.8091 & 0.8508 & 0.8443 & 0.7965 & 0.8030 \\
& L4 & High noise
& 1.0953 & \textbf{0.7905} & \underline{0.7911} & 0.8118 & 0.7919 & 0.8025 & 0.8126 & 0.8525 & 0.8472 & 0.7997 & 0.8034 \\
& L5 & Low SNR
& 1.1096 & \textbf{0.7846} & 0.7863 & 0.8075 & \underline{0.7848} & 0.7982 & 0.8056 & 0.8364 & 0.8197 & 0.7914 & 0.8007 \\
\midrule

\multirow{5}{*}{\textbf{3}} & L1 & Heavy tails
& 1.0999 & \textbf{0.7704} & \underline{0.7715} & 0.7839 & 0.7721 & 0.7837 & 0.7914 & 0.8417 & 0.8242 & 0.7808 & 0.7838 \\
& L2 & Extreme tails
& 0.8228 & \textbf{0.5634} & \underline{0.5651} & 0.5894 & 0.5657 & 0.5764 & 0.5909 & 0.6512 & 0.6384 & 0.5767 & 0.5792 \\
& L3 & Frequent outliers
& 1.0146 & \textbf{0.6996} & \underline{0.7011} & 0.7201 & 0.7020 & 0.7076 & 0.7167 & 0.7695 & 0.7610 & 0.7105 & 0.7087 \\
& L4 & Large outliers
& 0.9455 & \textbf{0.6474} & \underline{0.6482} & 0.6728 & 0.6498 & 0.6562 & 0.6626 & 0.7381 & 0.7250 & 0.6631 & 0.6558 \\
& L5 & Worst-case tails
& 0.7782 & \textbf{0.5036} & \underline{0.5059} & 0.5390 & 0.5087 & 0.5107 & 0.5243 & 0.6544 & 0.6269 & 0.5289 & 0.5120 \\
\midrule

\multirow{5}{*}{\textbf{4}} & L1 & Moderate regimes
& 1.0040 & \textbf{0.7839} & \underline{0.7845} & 0.8037 & 0.7910 & 0.7850 & 0.7861 & 0.8755 & 0.8565 & 0.8064 & 0.7848 \\
& L2 & Frequent switches
& 0.9992 & \textbf{0.7808} & \underline{0.7812} & 0.7992 & 0.7875 & 0.7850 & 0.7887 & 0.9037 & 0.8620 & 0.8017 & 0.7822 \\
& L3 & Subtle regimes
& 1.0068 & \textbf{0.7801} & \underline{0.7806} & 0.7961 & 0.7866 & 0.7871 & 0.7876 & 0.8568 & 0.8484 & 0.8015 & 0.7855 \\
& L4 & Strong regimes
& 0.9025 & \underline{0.7144} & \textbf{0.7126} & 0.7218 & 0.7176 & 0.7183 & 0.7174 & 0.7884 & 0.7909 & 0.7394 & 0.7152 \\
& L5 & Persistent regimes
& 0.7113 & \textbf{0.6357} & \underline{0.6365} & 0.6518 & 0.7497 & 0.6398 & 0.6424 & 0.8434 & 0.8271 & 0.7502 & 0.6408 \\
\midrule

\multirow{5}{*}{\textbf{5}} & L1 & Moderate clustering
& 1.0249 & \textbf{0.7481} & \underline{0.7490} & 0.7658 & 0.7505 & 0.7571 & 0.7628 & 0.8259 & 0.8034 & 0.7633 & 0.7577 \\
& L2 & Strong clustering
& 1.0641 & \textbf{0.7743} & \underline{0.7754} & 0.7905 & 0.7755 & 0.7819 & 0.7862 & 0.8444 & 0.8350 & 0.7878 & 0.7811 \\
& L3 & Long-memory clustering
& 1.0181 & \textbf{0.7364} & \underline{0.7386} & 0.7595 & 0.7414 & 0.7531 & 0.7648 & 0.8157 & 0.8050 & 0.7508 & 0.7530 \\
& L4 & High jump rate
& 1.0494 & \textbf{0.7626} & \underline{0.7631} & 0.7878 & 0.7654 & 0.7720 & 0.7812 & 0.8483 & 0.8302 & 0.7808 & 0.7730 \\
& L5 & Heavy-tailed jumps
& 0.8582 & \textbf{0.6018} & \underline{0.6031} & 0.6350 & 0.6059 & 0.6326 & 0.6304 & 0.7233 & 0.6758 & 0.6148 & 0.6256 \\
\midrule

\multirow{5}{*}{\textbf{6}} & L1 & Baseline
& 0.9243 & \textbf{0.6926} & \underline{0.6937} & 0.7095 & 0.7031 & 0.7056 & 0.7148 & 0.7911 & 0.7684 & 0.7151 & 0.7057 \\
& L2 & Rare events
& 0.9483 & \textbf{0.7215} & \underline{0.7223} & 0.7374 & 0.7319 & 0.7292 & 0.7334 & 0.8061 & 0.7930 & 0.7460 & 0.7287 \\
& L3 & Bursty events
& 0.8649 & \textbf{0.6334} & \underline{0.6364} & 0.6630 & 0.6521 & 0.6731 & 0.6764 & 0.7319 & 0.7125 & 0.6607 & 0.6618 \\
& L4 & Heavy-tailed events
& 0.7810 & \textbf{0.5764} & \underline{0.5785} & 0.7074 & 0.5962 & 0.6318 & 0.6154 & 0.7686 & 0.7011 & 0.6004 & 0.6088 \\
& L5 & Persistent background
& 0.6645 & \textbf{0.5618} & \underline{0.5627} & 0.5791 & 0.6726 & 0.5814 & 0.5838 & 0.7443 & 0.7148 & 0.6570 & 0.5804 \\
\bottomrule
\end{tabular}
}

\vspace{1mm}
\begin{minipage}{0.98\textwidth}
\footnotesize
\textbf{Abbrev:} iTrans=iTransformer; AutoF=Autoformer; FEDF=FEDformer; NTrans=Nonstationary Transformer.\\
\textbf{Note:} Although originally proposed for modeling realized volatility, HAR is applied here as a linear multi-scale autoregressive baseline for mean forecasting on the target series.
\end{minipage}

\end{table*}

\begin{figure*}[t]
  \centering
  \includegraphics[width=0.8\textwidth]{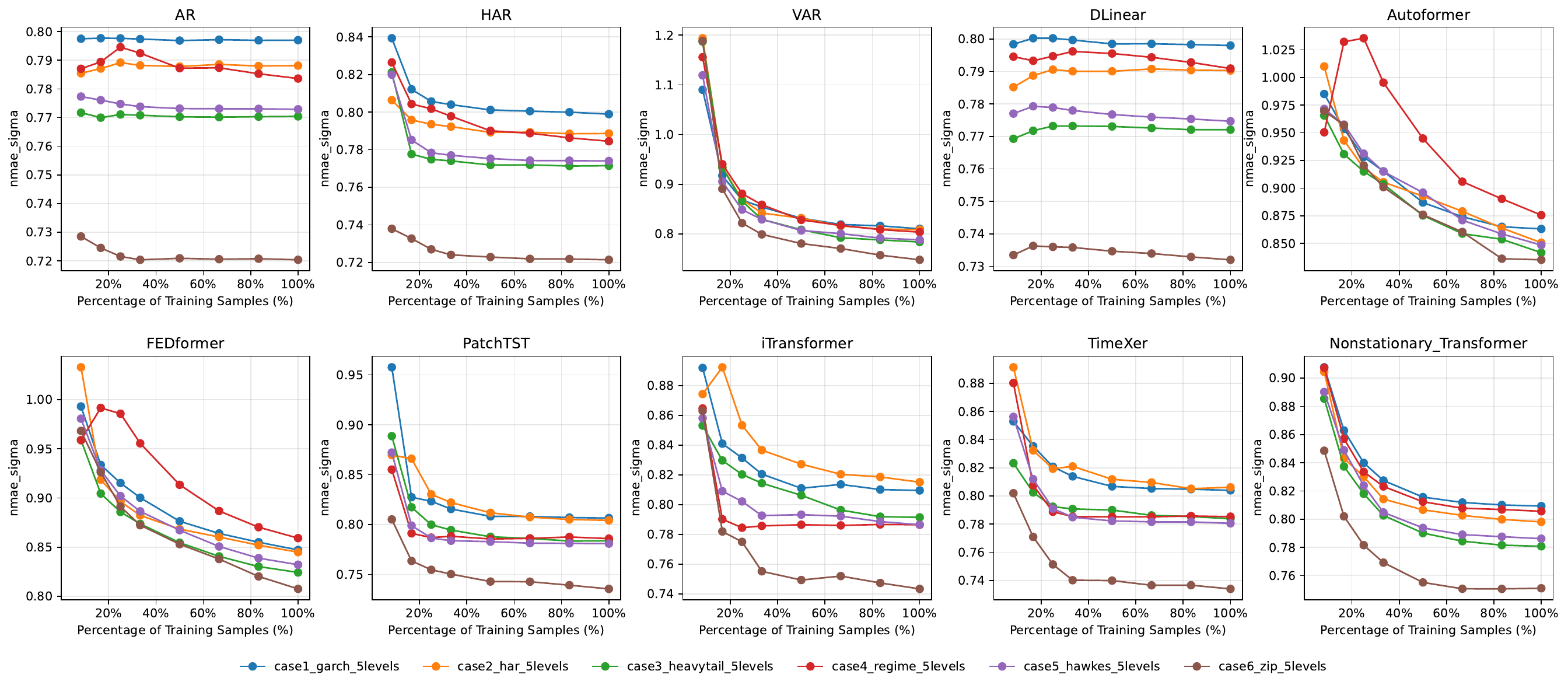}
  \caption{
Data-efficiency learning curves for 10 models across six synthetic mechanism families.
Each subplot shows $\mathrm{NMAE}_\sigma$ versus training set size.
}
\label{fig:data-efficiency}
\end{figure*}

\subsection{Evaluation Metrics}
\label{sec:metrics}

We evaluate point and probabilistic forecasts using scale-normalized metrics that are comparable across mechanism families and sensitive to uncertainty calibration.

\paragraph{\noindent \textbf{Point forecasting metric.}}
For point forecasters, we report a volatility-normalized mean absolute error, denoted as $\mathrm{NMAE}_\sigma$. Let $\hat{y}_{i,t}$ be the one-step-ahead prediction for series $i$ at time $t$, and $y_{i,t}$ the ground truth. Over the test window containing $T_{\text{test}}$ time steps and $N$ series, we define
\begin{equation}
\mathrm{NMAE}_\sigma
=
\frac{
\frac{1}{T_{\text{test}}N}
\sum_{i=1}^{N}\sum_{t=1}^{T_{\text{test}}}
\left|y_{i,t}-\hat{y}_{i,t}\right|
}{
\mathrm{Std}\!\left(\{y_{i,t}\}_{i=1..N,\;t=1..T_{\text{test}}}\right)
}.
\label{eq:nmae_sigma}
\end{equation}
The denominator is the pooled empirical standard deviation over all series and time steps in the test set. This measures error relative to the intrinsic volatility of the data, enabling fair comparison across mechanisms with different scales.

\paragraph{\noindent \textbf{Probabilistic forecasting metric.}}
For probabilistic forecasters, we use the Continuous Ranked Probability Score (CRPS)~\cite{matheson1976scoring}, which evaluates the agreement between a predictive cumulative distribution function $F$ and an observation $y$:
\begin{equation}
\mathrm{CRPS}(F,y)
=
\int_{\mathbb{R}}
\big(F(z)-\mathbf{1}\{y\le z\}\big)^2\,dz.
\label{eq:crps_def}
\end{equation}
CRPS is a proper scoring rule and is minimized in expectation when the predictive distribution matches the data-generating distribution.

In practice, probabilistic models provide Monte Carlo samples $\{x^{(s)}\}_{s=1}^{S}\sim F$. We compute CRPS using the sample-based estimator
\begin{equation}
\widehat{\mathrm{CRPS}}(F,y)
=
\frac{1}{S}\sum_{s=1}^{S}|x^{(s)}-y|
-
\frac{1}{2S^2}\sum_{s=1}^{S}\sum_{s'=1}^{S}|x^{(s)}-x^{(s')}|.
\label{eq:crps_sample}
\end{equation}

For multivariate time series, we evaluate calibration on the cross-sectional aggregate, reflecting portfolio-level distributional accuracy. Let
\(
y^{\mathrm{sum}}_{w,h}=\sum_{i=1}^{N} y_{i,w,h}
\)
and
\(
x^{\mathrm{sum}(s)}_{w,h}=\sum_{i=1}^{N} x^{(s)}_{i,w,h}
\)
denote the summed ground truth and sampled forecasts for evaluation window $w$ and horizon $h$. We compute
\begin{equation}
\mathrm{CRPS}_{\mathrm{sum}}
=
\frac{1}{WH}\sum_{w=1}^{W}\sum_{h=1}^{H}
\widehat{\mathrm{CRPS}}\!\left(F^{\mathrm{sum}}_{w,h},\,
y^{\mathrm{sum}}_{w,h}\right).
\label{eq:crps_sum}
\end{equation}
To compare across datasets with different scales, we report a normalized version by dividing $\mathrm{CRPS}_{\mathrm{sum}}$ by
\(
\frac{1}{WH}\sum_{w,h}|y^{\mathrm{sum}}_{w,h}|.
\)

\paragraph{\noindent \textbf{Data Efficiency.}} To quantify sample complexity, we train models on increasing training sizes
$n \in \{100, 200, 300, 400, 600, 800, 1000,\allowbreak 1200\}$ while keeping the test set fixed, constructing learning curves (Figure~\ref{fig:data-efficiency}) to identify the minimum data required for neural models to outperform linear baselines.

\begin{table*}[t] 
  \centering
  \begin{minipage}[b]{0.62\textwidth}
    \caption{Probabilistic forecasting results (CRPS $\downarrow$).}
    \vspace{-5px}
    \label{tab:main_results_prob}
    \scriptsize
    \setlength{\tabcolsep}{2.6pt}
    \renewcommand{\arraystretch}{1.05}
    \resizebox{\linewidth}{!}{%
      \begin{tabular}{lllcccccc}

\toprule
\multirow{2}{*}{\textbf{Case}} &
\multirow{2}{*}{\textbf{Level}} &
\multirow{2}{*}{\textbf{Scenario}} &
\multicolumn{6}{c}{\textbf{Probabilistic models (CRPS $\downarrow$)}} \\
\cmidrule(lr){4-9}
& & &
\textbf{DeepAR} &
\textbf{TimeGrad} &
\textbf{TSFlow} &
\textbf{TimeMCL} &
\textbf{RATD} &
\textbf{QFormer} \\
\midrule

\multirow{5}{*}{\textbf{1}} & L1 & Baseline
& \underline{0.7346} & 1.0687 & 1.0695 & 0.9667 & \textbf{0.7208} & 0.9604 \\
& L2 & Factor persistence
& \textbf{0.7378} & 1.0579 & 1.0296 & 0.9636 & \underline{0.7493} & 0.9697 \\
& L3 & Idio persistence
& \textbf{0.7371} & 1.0629 & \underline{0.8316} & 1.0166 & 0.9092 & 0.9558 \\
& L4 & Heterogeneity
& \textbf{0.7323} & 1.0698 & 0.9935 & 1.0597 & \underline{0.8290} & 0.9597 \\
& L5 & Low SNR
& \textbf{0.7438} & 1.0210 & 0.9748 & 1.0341 & 1.1623 & \underline{0.9394} \\
\midrule

\multirow{5}{*}{\textbf{2}} & L1 & Baseline
& \textbf{0.7312} & 1.0491 & 1.1959 & 0.9911 & \underline{0.7863} & 0.9951 \\
& L2 & High persistence
& \textbf{0.7316} & 1.0807 & 0.9587 & 1.0310 & \underline{0.7922} & 0.9920 \\
& L3 & Long memory
& \textbf{0.7281} & 1.1038 & 1.0526 & 1.0276 & \underline{0.7800} & 0.9955 \\
& L4 & High noise
& \textbf{0.7363} & 1.0826 & 0.9791 & 0.9760 & \underline{0.7906} & 0.9890 \\
& L5 & Low SNR
& \textbf{0.7338} & 1.0603 & 1.1436 & 1.0865 & \underline{0.7395} & 0.9952 \\
\midrule

\multirow{5}{*}{\textbf{3}} & L1 & Heavy tails
& \textbf{0.7370} & 1.0454 & 0.9929 & 1.0062 & \underline{0.8465} & 0.9583 \\
& L2 & Extreme tails
& \textbf{0.7570} & 1.0052 & 1.0136 & 0.9653 & \underline{0.9188} & 0.9647 \\
& L3 & Frequent outliers
& \textbf{0.7628} & 1.0832 & 1.1309 & 0.9932 & 0.9643 & \underline{0.9503} \\
& L4 & Large outliers
& \textbf{0.7618} & 1.0372 & \underline{0.8987} & 1.0314 & 0.9674 & 0.9497 \\
& L5 & Worst-case tails
& \textbf{0.7654} & 0.9573 & \underline{0.8541} & 1.0564 & 1.2393 & 0.9457 \\
\midrule

\multirow{5}{*}{\textbf{4}} & L1 & Moderate regimes
& \underline{0.6421} & 0.6840 & \textbf{0.6039} & 0.8521 & 1.9729 & 0.9902 \\
& L2 & Frequent switches
& \textbf{0.6628} & 0.7951 & \underline{0.7809} & 0.9244 & 2.2090 & 0.9926 \\
& L3 & Subtle regimes
& \textbf{0.8121} & 0.9383 & \underline{0.8409} & 1.0200 & 2.2031 & 0.9919 \\
& L4 & Strong regimes
& \underline{0.9844} & \textbf{0.9709} & 1.3420 & 1.5561 & 1.2589 & 0.9970 \\
& L5 & Persistent regimes
& \textbf{0.5333} & \underline{0.6836} & 0.7799 & 0.7767 & 1.9097 & 0.9828 \\
\midrule

\multirow{5}{*}{\textbf{5}} & L1 & Moderate clustering
& \textbf{0.9051} & 1.1326 & 1.1165 & 1.1689 & 1.4851 & \underline{0.9969} \\
& L2 & Strong clustering
& \textbf{0.9095} & 1.2770 & 1.3267 & 1.1077 & 1.4627 & \underline{0.9960} \\
& L3 & Long-memory clustering
& \textbf{0.9338} & 1.2030 & 1.2134 & 1.1073 & 1.2756 & \underline{0.9968} \\
& L4 & High jump rate
& \textbf{0.9262} & 1.2509 & 1.5418 & 1.1599 & 1.2673 & \underline{0.9964} \\
& L5 & Heavy-tailed jumps
& \textbf{0.9727} & 1.1350 & 1.7447 & 1.3144 & 1.8884 & \underline{0.9961} \\
\midrule

\multirow{5}{*}{\textbf{6}} & L1 & Baseline
& \textbf{0.9270} & 1.1704 & 1.1310 & 1.0982 & 1.4757 & \underline{0.9951} \\
& L2 & Rare events
& \underline{0.8769} & 1.1403 & \textbf{0.8379} & 1.0434 & 1.7746 & 0.9945 \\
& L3 & Bursty events
& \textbf{0.9724} & 1.0791 & 1.2106 & 1.0984 & 1.2565 & \underline{0.9977} \\
& L4 & Heavy-tailed events
& 1.0548 & 1.1237 & \textbf{0.8055} & 1.6101 & 2.3690 & \underline{0.9940} \\
& L5 & Persistent background
& \underline{0.9157} & 1.0738 & \textbf{0.8100} & 1.0692 & 1.2109 & 0.9906 \\

\bottomrule
\end{tabular}
    }
    \vspace{2mm}
    \begin{flushleft}
    \footnotesize
    \textbf{Abbrev:} QFormer=QuantileFormer.
    \end{flushleft}
  \end{minipage}
  \hfill 
  \begin{minipage}[b]{0.36\textwidth}
    \centering
    \includegraphics[width=\linewidth]{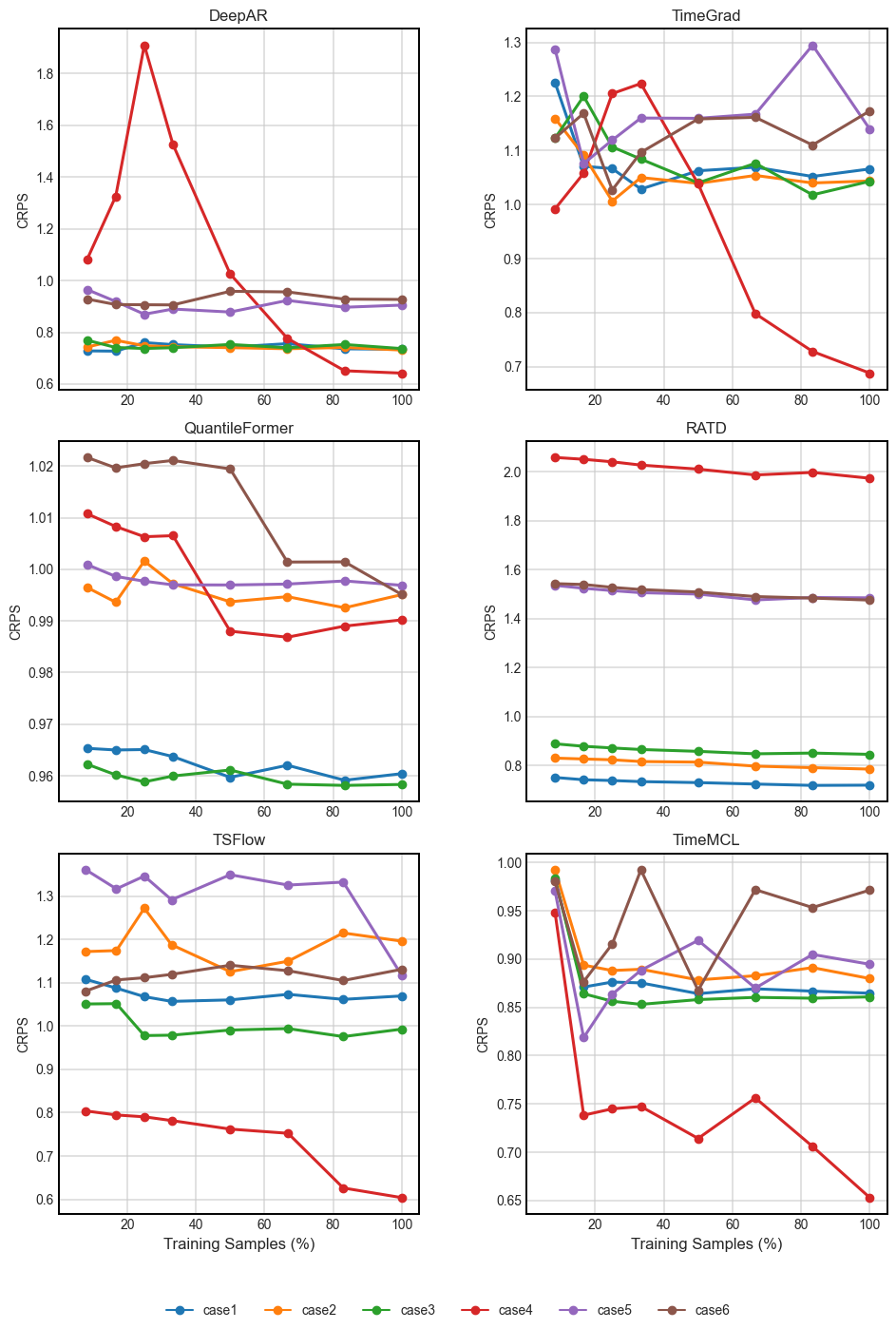}
    \captionof{figure}{CRPS as a function of training sample ratio across the six synthetic cases for each probabilistic model. Each line corresponds to one case (Case 1--6). Lower is better.}
    \label{fig:crps_train_ratio}
  \end{minipage}
\end{table*}
\section{Results}
\label{sec:results}

\subsection{Point Forecasting: The Dominance of Robustness}
\label{sec:results_point}

Table~\ref{tab:main_results_point} reports NMAE$_\sigma$ across all environments. Within this controlled setting, the results suggest that model performance is more closely associated with mechanism-compatible inductive biases than with model capacity alone.

\paragraph{\noindent \textbf{Finding 1: Robustness trumps expressiveness.}} Simple baselines (AR, HAR) and the linear decomposition model (DLinear) consistently outperform complex Transformers. This indicates that in low signal-to-noise financial environments (Cases 1--3), the primary task is robust mean-reversion rather than complex pattern matching. Deep models with high capacity (e.g., Nonstationary Transformer) tend to overfit idiosyncratic noise, whereas DLinear's structural constraint---modeling trends without attention---acts as a beneficial regularizer against overreaction to jumps and outliers.


\paragraph{\noindent \textbf{Finding 2: Local attention outperforms global mixing.}} Among attention-based models, PatchTST typically ranks highest. Its patch-based design preserves local temporal structure without forcing global cross-series mixing. In contrast, architectures designed for global correlation (iTransformer, TimeXer) struggle in jump-driven environments (Cases 5--6). This suggests that global attention mechanisms may diffuse the impact of local shocks, diluting the signal required to predict sudden moves.


\paragraph{\textbf{Finding 3: Decomposition fails without periodicity.}}
Autoformer and FEDformer, which rely on seasonal–trend decomposition and frequency-domain representations, underperform consistently across nearly all synthetic environments.
Unlike traffic, energy, or weather data, financial return series tend to exhibit weaker periodic structure and are dominated instead by stochastic volatility, regime changes, and heavy-tailed shocks.
As a result, inductive biases designed to extract periodic components provide little advantage and can even hinder performance.
These results highlight a fundamental mismatch between popular time-series inductive biases and the structural properties of financial data. Notably, many of these architectures were originally designed for long-horizon forecasting in strongly periodic domains (e.g., traffic, energy, weather), whereas our evaluation focuses on one-step-ahead prediction in noise-dominated environments. This contrast further illustrates that inductive biases effective for long-term seasonal prediction do not necessarily transfer to short-horizon financial forecasting.

\subsection{Probabilistic Forecasting: The Role of Inductive Bias}
\label{sec:results_prob}

Table~\ref{tab:main_results_prob} reports CRPS, revealing how well a model's distributional assumptions align with the mechanism.

\paragraph{\noindent \textbf{Finding 4: Parametric alignment yields efficiency.}}
DeepAR achieves the best CRPS in 24 of 30 settings. Its success is not accidental: its autoregressive Gaussian likelihood with time-varying scale ($\sigma_t$) is structurally isomorphic to the GARCH dynamics generating Case 1 and Case 2.
This confirms that \emph{correct parametric specification} (even if simple) beats flexible but data-hungry density estimators in stationary regimes.


\paragraph{\noindent \textbf{Finding 5: Flexibility wins under multimodality.}}
The limits of DeepAR are exposed in Case~4 (Regime Switching) and Case~6 (Zero-Inflated Jumps).
Here, the true posterior is multimodal (mixture of regimes) or zero-inflated.
TSFlow, which uses normalizing flows to model arbitrary densities, overtakes DeepAR in these settings (e.g., Case~4~L1), proving that generative flexibility is essential when the unimodal Gaussian assumption is violated.

\paragraph{\noindent \textbf{Finding 6: Diffusion models struggle with structural breaks.}}
RATD performs well on smooth volatility processes but degrades on discontinuous mechanisms (Cases 4 \& 6).
This suggests that diffusion-based residuals struggle to adapt to abrupt structural breaks where the "noise" distribution shifts instantaneously.

\subsection{Data Efficiency: The Cost of Complexity}
\label{sec:results_point_dataeff}

Figure~\ref{fig:data-efficiency} (Point Forecasting) reveals a stark efficiency gap between statistical and neural methods.

\paragraph{\noindent \textbf{The ``Early Saturation'' Effect.}}
For volatility (Case 1) and tail-driven (Case 3) mechanisms, performance saturates rapidly.
Neural models improve up to $\sim$40\% of the training data and then plateau.
This implies that the "signal" in one-step financial forecasting is information-sparse; providing more data effectively just samples more noise, offering diminishing returns for complex architectures.

\paragraph{\noindent \textbf{Regimes demand data.}}
The only exception is Case 4 (Regime Switching), where learning curves remain steep even at larger sample sizes.
This supports that inferring latent structural changes is an information-dense task requiring significantly more history than learning stationary volatility dynamics.

\subsection{Probabilistic Data Efficiency}
\label{sec:results_prob_dataeff}

Figure 3 (Probabilistic CRPS) highlights that calibration is far more data-intensive than point prediction.

\paragraph{\noindent \textbf{Data-insensitive baselines.}}
DeepAR is remarkably data-efficient in stationary settings (Cases 1--3), stabilizing with very few samples.
This robustness makes it a strong baseline for data-scarce financial applications, provided the stationarity assumption holds.

\paragraph{\noindent \textbf{Generative models require scale.}}
Flexible models like TSFlow and TimeMCL show visible improvements as data volume increases, particularly in complex Case 4.
Unlike DeepAR, they must learn the distributional shape from scratch (without a parametric prior).
Consequently, they require 2--3$\times$ more data to achieve competitive calibration, highlighting a clear trade-off between \emph{asymptotic flexibility} and \emph{small-sample robustness}.
\section{Limitations and Future Directions}
\label{sec:limitations}

FinStressTS is a diagnostic benchmark rather than a full market simulator, and we note several limitations that motivate future extensions. First, as a parametric simulator, \bench{} does not model many real-market complexities (e.g., limit order books). Thus, strong performance should be interpreted as robustness and calibration \emph{under specified mechanisms}, not as a guarantee of trading profitability. We view \bench{} as complementary to real-data evaluation. Second, each of the six mechanism families is implemented via canonical econometric formulations. These choices are well-grounded but not exhaustive; alternative DGPs (e.g., stochastic volatility) and parameter configurations may induce different learning dynamics and shift relative rankings. An important next step is to add multiple variants per mechanism family and test robustness across formulations. Third, our benchmark focuses on fixed-horizon forecasting and does not yet cover (i) multi-step forecasting, (ii) online adaptation under nonstationarity, or (iii) decision-aware evaluation for downstream tasks (portfolio allocation). While \bench{} supports compound stresses, expanding interacting mechanism combinations (e.g., regime shifts with heavy tails and jumps) would further enrich diagnostics. Last, although we evaluate 15 diverse models, the space of time-series methods is evolving, requiring ongoing benchmark updates. Learning curves and distributional metrics also add computational cost, which may limit diagnostic granularity for very large models. We mitigate this with standardized protocols and release \bench{} as an extensible artifact to support community-driven expansion.
\section{Conclusion}
\label{sec:conclusion}

We introduced \bench, a mechanism-aware synthetic benchmark that addresses a fundamental gap in financial forecasting evaluation: the inability to isolate \emph{why} models fail. By organizing 30 diagnostic environments around six canonical mechanisms with parametric control, \bench{} transforms opaque failures into interpretable diagnoses, enabling researchers to trace underperformance to precise structural causes rather than confounded explanations inherent in historical data.

Our evaluation of 15 models across point and probabilistic tasks reveals four insights impossible to obtain from black-box benchmarks. \textit{First}, simple autoregressive models dominate one-step mean forecasting across nearly all mechanisms, showing that financial prediction difficulty lies primarily in modeling uncertainty, not conditional means. \textit{Second}, performance is highly mechanism-dependent: models excelling under stationary volatility (e.g., DeepAR) degrade catastrophically under regime switching, where flexible density models (e.g., TSFlow) gain advantage. \textit{Third}, ground-truth access via known DGPs reveals persistent probabilistic miscalibration—deep models fail at reliable uncertainty quantification. \textit{Fourth}, learning curves expose stark efficiency gaps: neural methods require 2--3$\times$ more samples than classical baselines for comparable calibration, with some never reaching parity—particularly concerning given data scarcity from regime shifts in finance.

These findings have direct implications: when data are limited or regimes shift, sophisticated neural architectures may underperform simpler alternatives. Mechanism-aware model selection matters more than universal sophistication; probabilistic calibration must be validated independently from point accuracy; and data efficiency should be a first-order design consideration. Beyond empirical insights, \bench{} provides open-source infrastructure enabling researchers to stress-test architectures, validate robustness claims, and systematically compare models under controlled conditions—a living resource the community can extend with additional mechanisms, compound scenarios, and multi-step evaluation.



\bibliographystyle{ACM-Reference-Format}
\balance
\bibliography{citations}

@article{makridakis2024m6,
  title={The M6 forecasting competition: Bridging the gap between forecasting and investment decisions},
  author={Makridakis, Spyros and Spiliotis, Evangelos and Hollyman, Ross and Petropoulos, Fotios and Swanson, Norman and Gaba, Anil},
  journal={International Journal of Forecasting},
  year={2024},
  publisher={Elsevier BV}
}

@article{cont2001stylized,
  title   = {Empirical properties of asset returns: stylized facts and statistical issues},
  author  = {Cont, Rama},
  journal = {Quantitative Finance},
  volume  = {1},
  number  = {2},
  pages   = {223--236},
  year    = {2001},
  doi     = {10.1080/713665670}
}

@article{engle1982arch,
  title   = {Autoregressive Conditional Heteroskedasticity with Estimates of the Variance of United Kingdom Inflation},
  author  = {Engle, Robert F.},
  journal = {Econometrica},
  volume  = {50},
  number  = {4},
  pages   = {987--1007},
  year    = {1982},
  doi     = {10.2307/1912773}
}

@article{bollerslev1986garch,
  title   = {Generalized Autoregressive Conditional Heteroskedasticity},
  author  = {Bollerslev, Tim},
  journal = {Journal of Econometrics},
  volume  = {31},
  number  = {3},
  pages   = {307--327},
  year    = {1986},
  doi     = {10.1016/0304-4076(86)90063-1}
}

@article{corsi2009har,
  title   = {A Simple Approximate Long-Memory Model of Realized Volatility},
  author  = {Corsi, Fulvio},
  journal = {Journal of Financial Econometrics},
  volume  = {7},
  number  = {2},
  pages   = {174--196},
  year    = {2009},
  doi     = {10.1093/jjfinec/nbp001}
}

@article{bollerslev1987garch,
  title   = {A Conditionally Heteroskedastic Time Series Model for Speculative Prices and Rates of Return},
  author  = {Bollerslev, Tim},
  journal = {The Review of Economics and Statistics},
  volume  = {69},
  number  = {3},
  pages   = {542--547},
  year    = {1987}
}

@article{hawkes1971hawkes,
  title   = {Spectra of some self-exciting and mutually exciting point processes},
  author  = {Hawkes, Alan G.},
  journal = {Biometrika},
  volume  = {58},
  number  = {1},
  pages   = {83--90},
  year    = {1971},
  doi     = {10.1093/biomet/58.1.83}
}

@article{lambert1992zip,
  title   = {Zero-Inflated Poisson Regression, with an Application to Defects in Manufacturing},
  author  = {Lambert, Diane},
  journal = {Technometrics},
  volume  = {34},
  number  = {1},
  pages   = {1--14},
  year    = {1992},
  doi     = {10.1080/00401706.1992.10485228}
}

@article{gneiting2007scoring,
  title   = {Strictly Proper Scoring Rules, Prediction, and Estimation},
  author  = {Gneiting, Tilmann and Raftery, Adrian E.},
  journal = {Journal of the American Statistical Association},
  volume  = {102},
  number  = {477},
  pages   = {359--378},
  year    = {2007},
  doi     = {10.1198/016214506000001437}
}

@article{andersen2003rv,
  title   = {Modeling and Forecasting Realized Volatility},
  author  = {Andersen, Torben G. and Bollerslev, Tim and Diebold, Francis X. and Labys, Paul},
  journal = {Econometrica},
  volume  = {71},
  number  = {2},
  pages   = {579--625},
  year    = {2003},
  doi     = {10.1111/1468-0262.00402}
}

@book{embrechts1997extremes,
  title     = {Modelling Extremal Events for Insurance and Finance},
  author    = {Embrechts, Paul and Kl{\"u}ppelberg, Claudia and Mikosch, Thomas},
  publisher = {Springer},
  year      = {1997},
  doi       = {10.1007/978-3-642-33483-2}
}

@article{bacry2015hawkes,
  title   = {Hawkes Processes in Finance},
  author  = {Bacry, Emmanuel and Mastromatteo, Iacopo and Muzy, Jean-Fran{\c{c}}ois},
  journal = {Market Microstructure and Liquidity},
  volume  = {1},
  number  = {1},
  pages   = {1550005},
  year    = {2015},
  doi     = {10.1142/S2382626615500057}
}

@article{lesmond1999costs,
  title   = {A New Estimate of Transaction Costs},
  author  = {Lesmond, David A. and Ogden, Joseph P. and Trzcinka, Charles A.},
  journal = {The Review of Financial Studies},
  volume  = {12},
  number  = {5},
  pages   = {1113--1141},
  year    = {1999},
  doi     = {10.1093/rfs/12.5.1113}
}

@article{box1976analysis,
  title={Analysis: Forecasting and Control},
  author={Box, George and Jenkins, GM},
  journal={San francisco},
  year={1976}
}

@book{lutkepohl2013introduction,
  title={Introduction to multiple time series analysis},
  author={L{\"u}tkepohl, Helmut},
  year={2013},
  publisher={Springer Science \& Business Media}
}

@article{corsi2009simple,
  title={A simple approximate long-memory model of realized volatility},
  author={Corsi, Fulvio},
  journal={Journal of financial econometrics},
  volume={7},
  number={2},
  pages={174--196},
  year={2009},
  publisher={Oxford University Press}
}

@article{andersen2003modeling,
  title={Modeling and forecasting realized volatility},
  author={Andersen, Torben G and Bollerslev, Tim and Diebold, Francis X and Labys, Paul},
  journal={Econometrica},
  volume={71},
  number={2},
  pages={579--625},
  year={2003},
  publisher={Wiley Online Library}
}

@article{cont2001empirical,
  title={Empirical properties of asset returns: stylized facts and statistical issues},
  author={Cont, Rama},
  journal={Quantitative finance},
  volume={1},
  number={2},
  pages={223},
  year={2001},
  publisher={IOP Publishing}
}

@article{makridakis2018m4,
  title={The M4 Competition: Results, findings, conclusion and way forward},
  author={Makridakis, Spyros and Spiliotis, Evangelos and Assimakopoulos, Vassilios},
  journal={International Journal of forecasting},
  volume={34},
  number={4},
  pages={802--808},
  year={2018},
  publisher={Elsevier}
}

@inproceedings{zeng2023transformers,
  title={Are transformers effective for time series forecasting?},
  author={Zeng, Ailing and Chen, Muxi and Zhang, Lei and Xu, Qiang},
  booktitle={Proceedings of the AAAI conference on artificial intelligence},
  volume={37},
  number={9},
  pages={11121--11128},
  year={2023}
}

@inproceedings{zhou2022fedformer,
  title={Fedformer: Frequency enhanced decomposed transformer for long-term series forecasting},
  author={Zhou, Tian and Ma, Ziqing and Wen, Qingsong and Wang, Xue and Sun, Liang and Jin, Rong},
  booktitle={International conference on machine learning},
  pages={27268--27286},
  year={2022},
  organization={PMLR}
}

@article{wu2021autoformer,
  title={Autoformer: Decomposition transformers with auto-correlation for long-term series forecasting},
  author={Wu, Haixu and Xu, Jiehui and Wang, Jianmin and Long, Mingsheng},
  journal={Advances in neural information processing systems},
  volume={34},
  pages={22419--22430},
  year={2021}
}

@article{nie2022time,
  title={A Time Series is Worth 64Words: Long-term Forecasting with Transformers},
  author={Nie, Y},
  journal={arXiv preprint arXiv:2211.14730},
  year={2022}
}

@article{liu2023itransformer,
  title={itransformer: Inverted transformers are effective for time series forecasting},
  author={Liu, Yong and Hu, Tengge and Zhang, Haoran and Wu, Haixu and Wang, Shiyu and Ma, Lintao and Long, Mingsheng},
  journal={arXiv preprint arXiv:2310.06625},
  year={2023}
}

@article{wang2024timexer,
  title={Timexer: Empowering transformers for time series forecasting with exogenous variables},
  author={Wang, Yuxuan and Wu, Haixu and Dong, Jiaxiang and Qin, Guo and Zhang, Haoran and Liu, Yong and Qiu, Yunzhong and Wang, Jianmin and Long, Mingsheng},
  journal={Advances in Neural Information Processing Systems},
  volume={37},
  pages={469--498},
  year={2024}
}

@article{liu2022non,
  title={Non-stationary transformers: Exploring the stationarity in time series forecasting},
  author={Liu, Yong and Wu, Haixu and Wang, Jianmin and Long, Mingsheng},
  journal={Advances in neural information processing systems},
  volume={35},
  pages={9881--9893},
  year={2022}
}

@article{gneiting2014probabilistic,
  title={Probabilistic forecasting},
  author={Gneiting, Tilmann and Katzfuss, Matthias},
  journal={Annual Review of Statistics and Its Application},
  volume={1},
  number={1},
  pages={125--151},
  year={2014},
  publisher={Annual Reviews}
}

@article{salinas2020deepar,
  title={DeepAR: Probabilistic forecasting with autoregressive recurrent networks},
  author={Salinas, David and Flunkert, Valentin and Gasthaus, Jan and Januschowski, Tim},
  journal={International journal of forecasting},
  volume={36},
  number={3},
  pages={1181--1191},
  year={2020},
  publisher={Elsevier}
}

@article{alexandrov2020gluonts,
  title={Gluonts: Probabilistic and neural time series modeling in python},
  author={Alexandrov, Alexander and Benidis, Konstantinos and Bohlke-Schneider, Michael and Flunkert, Valentin and Gasthaus, Jan and Januschowski, Tim and Maddix, Danielle C and Rangapuram, Syama and Salinas, David and Schulz, Jasper and others},
  journal={Journal of Machine Learning Research},
  volume={21},
  number={116},
  pages={1--6},
  year={2020}
}

@article{kollovieh2024flow,
  title={Flow matching with gaussian process priors for probabilistic time series forecasting},
  author={Kollovieh, Marcel and Lienen, Marten and L{\"u}dke, David and Schwinn, Leo and G{\"u}nnemann, Stephan},
  journal={arXiv preprint arXiv:2410.03024},
  year={2024}
}

@inproceedings{cortes2025winner,
  title={Winner-takes-all for Multivariate Probabilistic Time Series Forecasting},
  author={Cort{\'e}s, Adrien and Rehm, R{\'e}mi and Letzelter, Victor},
  booktitle={ICML 2025: The 42nd International Conference on Machine Learning},
  year={2025}
}

@article{liu2024retrieval,
  title={Retrieval-augmented diffusion models for time series forecasting},
  author={Liu, Jingwei and Yang, Ling and Li, Hongyan and Hong, Shenda},
  journal={Advances in Neural Information Processing Systems},
  volume={37},
  pages={2766--2786},
  year={2024}
}

@inproceedings{shao2025quantileformer,
  title={QuantileFormer: Probabilistic Time Series Forecasting with a Pattern-Mixture Decomposed VAE Transformer},
  author={Shao, Yimiao and Li, Wenzhong and Xia, Kang and Lin, Kaijie and Lin, Mingkai and Lu, Sanglu},
  booktitle={Proceedings of the Thirty-Fourth International Joint Conference on Artificial Intelligence},
  pages={6147--6155},
  year={2025}
}

@article{godahewa2021monash,
  title={Monash time series forecasting archive},
  author={Godahewa, Rakshitha and Bergmeir, Christoph and Webb, Geoffrey I and Hyndman, Rob J and Montero-Manso, Pablo},
  journal={arXiv preprint arXiv:2105.06643},
  year={2021}
}

@article{taylor2018forecasting,
  title={Forecasting at scale},
  author={Taylor, Sean J and Letham, Benjamin},
  journal={The American Statistician},
  volume={72},
  number={1},
  pages={37--45},
  year={2018},
  publisher={Taylor \& Francis}
}

@article{loning2019sktime,
  title={sktime: A unified interface for machine learning with time series},
  author={L{\"o}ning, Markus and Bagnall, Anthony and Ganesh, Sajaysurya and Kazakov, Viktor and Lines, Jason and Kir{\'a}ly, Franz J},
  journal={arXiv preprint arXiv:1909.07872},
  year={2019}
}

@article{wang2024tssurvey,
  title={Deep Time Series Models: A Comprehensive Survey and Benchmark},
  author={Yuxuan Wang and Haixu Wu and Jiaxiang Dong and Yong Liu and Mingsheng Long and Jianmin Wang},
  booktitle={arXiv preprint arXiv:2407.13278},
  year={2024},
}

@misc{olivares2022library_neuralforecast,
    author={Kin G. Olivares and
            Cristian Challú and
            Azul Garza and
            Max Mergenthaler Canseco and
            Artur Dubrawski},
    title = {{NeuralForecast}: User friendly state-of-the-art neural forecasting models.},
    year={2022},
    howpublished={{PyCon} Salt Lake City, Utah, US 2022},
    url={https://github.com/Nixtla/neuralforecast}
}

@article{zhang2024probts,
  title={ProbTS: Benchmarking point and distributional forecasting across diverse prediction horizons},
  author={Zhang, Jiawen and Wen, Xumeng and Zhang, Zhenwei and Zheng, Shun and Li, Jia and Bian, Jiang},
  journal={Advances in Neural Information Processing Systems},
  volume={37},
  pages={48045--48082},
  year={2024}
}

@article{makridakis2000m3,
  title={The M3-Competition: results, conclusions and implications},
  author={Makridakis, Spyros and Hibon, Michele},
  journal={International journal of forecasting},
  volume={16},
  number={4},
  pages={451--476},
  year={2000},
  publisher={Elsevier}
}

@article{shao2024exploring,
 title={Exploring progress in multivariate time series forecasting: Comprehensive benchmarking and heterogeneity analysis},
 author={Shao, Zezhi and Wang, Fei and Xu, Yongjun and Wei, Wei and Yu, Chengqing and Zhang, Zhao and Yao, Di and Sun, Tao and Jin, Guangyin and Cao, Xin and others},
 journal={IEEE Transactions on Knowledge and Data Engineering},
 year={2024},
 volume={37},
 number={1},
 pages={291-305},
 publisher={IEEE}
}

@article{qiu2024tfb,
  title={TFB: Towards Comprehensive and Fair Benchmarking of Time Series Forecasting Methods},
  author={Qiu, Xiangfei and Hu, Jilin and Zhou, Lekui and Wu, Xingjian and Du, Junyang and Zhang, Buang and Guo, Chenjuan and Zhou, Aoying and Jensen, Christian S and Sheng, Zhenli and others},
  journal={Proceedings of the VLDB Endowment},
  volume={17},
  number={9},
  pages={2363--2377},
  year={2024},
  publisher={VLDB Endowment}
}

@inproceedings{wang2022adaptive,
  title={Adaptive Long-Short Pattern Transformer for Stock Investment Selection.},
  author={Wang, Heyuan and Wang, Tengjiao and Li, Shun and Zheng, Jiayi and Guan, Shijie and Chen, Wei},
  booktitle={IJCAI},
  pages={3970--3977},
  year={2022}
}

@inproceedings{chen2024automatic,
  title={Automatic de-biased temporal-relational modeling for stock investment recommendation},
  author={Chen, Weijun and Li, Shun and Yu, Xipu and Wang, Heyuan and Chen, Wei and Wang, Tengjiao},
  booktitle={Proceedings of the Thirty-Third International Joint Conference on Artificial Intelligence},
  pages={1999--2008},
  year={2024}
}

@inproceedings{duan2022factorvae,
  title={Factorvae: A probabilistic dynamic factor model based on variational autoencoder for predicting cross-sectional stock returns},
  author={Duan, Yitong and Wang, Lei and Zhang, Qizhong and Li, Jian},
  booktitle={Proceedings of the AAAI conference on artificial intelligence},
  volume={36},
  number={4},
  pages={4468--4476},
  year={2022}
}

@inproceedings{zeng2024trade,
  title={Trade when opportunity comes: price movement forecasting via locality-aware attention and iterative refinement labeling},
  author={Zeng, Liang and Wang, Lei and Niu, Hui and Zhang, Ruchen and Wang, Ling and Li, Jian},
  booktitle={Proceedings of the Thirty-Third International Joint Conference on Artificial Intelligence},
  pages={6134--6142},
  year={2024}
}

@article{hu2025fintsb,
  title={Fintsb: A comprehensive and practical benchmark for financial time series forecasting},
  author={Hu, Yifan and Li, Yuante and Liu, Peiyuan and Zhu, Yuxia and Li, Naiqi and Dai, Tao and Xia, Shu-tao and Cheng, Dawei and Jiang, Changjun},
  journal={arXiv preprint arXiv:2502.18834},
  year={2025}
}

@article{matheson1976scoring,
  title={Scoring rules for continuous probability distributions},
  author={Matheson, James E and Winkler, Robert L},
  journal={Management science},
  volume={22},
  number={10},
  pages={1087--1096},
  year={1976},
  publisher={INFORMS}
}

@article{kang2020gratis,
  title={GRATIS: GeneRAting TIme Series with diverse and controllable characteristics},
  author={Kang, Yanfei and Hyndman, Rob J and Li, Feng},
  journal={Statistical Analysis and Data Mining: The ASA Data Science Journal},
  volume={13},
  number={4},
  pages={354--376},
  year={2020},
  publisher={Wiley Online Library}
}

@article{nikitin2024tsgm,
  title={TSGM: a flexible framework for generative modeling of synthetic time series},
  author={Nikitin, Alexander and Iannucci, Letizia and Kaski, Samuel},
  journal={Advances in Neural Information Processing Systems},
  volume={37},
  pages={129042--129061},
  year={2024}
}

@article{ang2023tsgbench,
  title={TSGBench: Time Series Generation Benchmark},
  author={Ang, Yihao and Huang, Qiang and Bao, Yifan and Tung, Anthony KH and Huang, Zhiyong},
  journal={Proceedings of the VLDB Endowment},
  volume={17},
  number={3},
  pages={305--318},
  year={2023},
  publisher={VLDB Endowment}
}

@inproceedings{ferdous2025timegraph,
  title={Timegraph: Synthetic benchmark datasets for robust time-series causal discovery},
  author={Ferdous, Muhammad Hasan and Hossain, Emam and Gani, Md Osman},
  booktitle={Proceedings of the 31st ACM SIGKDD Conference on Knowledge Discovery and Data Mining V. 2},
  pages={5425--5435},
  year={2025}
}

@inproceedings{ozturk2024enhancing,
  title={Enhancing Financial Time-Series Analysis with TimeGAN: A Novel Approach},
  author={{\"O}zt{\"u}rk, Cemal},
  booktitle={2024 9th International Conference on Computer Science and Engineering (UBMK)},
  pages={447--450},
  year={2024},
  organization={IEEE}
}

@article{wiese2020quant,
  title={Quant GANs: deep generation of financial time series},
  author={Wiese, Magnus and Knobloch, Robert and Korn, Ralf and Kretschmer, Peter},
  journal={Quantitative Finance},
  volume={20},
  number={9},
  pages={1419--1440},
  year={2020},
  publisher={Taylor \& Francis}
}

@article{engle1982autoregressive,
  title={Autoregressive conditional heteroscedasticity with estimates of the variance of United Kingdom inflation},
  author={Engle, Robert F},
  journal={Econometrica: Journal of the econometric society},
  pages={987--1007},
  year={1982},
  publisher={JSTOR}
}

@article{hamilton1989new,
  title={A new approach to the economic analysis of nonstationary time series and the business cycle},
  author={Hamilton, James D},
  journal={Econometrica: Journal of the econometric society},
  pages={357--384},
  year={1989},
  publisher={JSTOR}
}

@article{andersen2007roughing,
  title={Roughing it up: Including jump components in the measurement, modeling, and forecasting of return volatility},
  author={Andersen, Torben G and Bollerslev, Tim and Diebold, Francis X},
  journal={The review of economics and statistics},
  volume={89},
  number={4},
  pages={701--720},
  year={2007},
  publisher={The MIT Press}
}

@article{fama1993common,
  title={Common risk factors in the returns on stocks and bonds},
  author={Fama, Eugene F and French, Kenneth R},
  journal={Journal of financial economics},
  volume={33},
  number={1},
  pages={3--56},
  year={1993},
  publisher={Elsevier}
}

@incollection{ross2013arbitrage,
  title={The arbitrage theory of capital asset pricing},
  author={Ross, Stephen A},
  booktitle={Handbook of the fundamentals of financial decision making: Part I},
  pages={11--30},
  year={2013},
  publisher={World Scientific}
}

@inproceedings{rasul2021autoregressive,
  title={Autoregressive denoising diffusion models for multivariate probabilistic time series forecasting},
  author={Rasul, Kashif and Seward, Calvin and Schuster, Ingmar and Vollgraf, Roland},
  booktitle={International conference on machine learning},
  pages={8857--8868},
  year={2021},
  organization={PMLR}
}

\appendix

\appendix

\section{Data Generation}
\label{sec:appendix_dgp}

FinStressTS generates multivariate panels of $N$ return series over $T$ retained time steps after burn-in. 
Each case targets one canonical financial mechanism, and its five diagnostic settings vary interpretable mechanism controls rather than defining a monotone difficulty scale.

\paragraph{Shared panel structure.}
For factor-based cases, returns are generated from a linear factor panel:
\begin{equation}
y_{i,t}
=
\alpha_i+\beta_i^\top f_t+u_{i,t},
\label{eq:shared_factor_panel}
\end{equation}
where $f_t\in\mathbb{R}^K$ denotes systematic factors, $u_{i,t}$ is the idiosyncratic component, and $(\alpha_i,\beta_i)$ are firm-specific intercepts and factor loadings drawn once per panel and fixed over time.
Conditional on the shared factors and variance paths, idiosyncratic residuals are independent across firms.
Thus, cross-sectional dependence is induced through common latent drivers rather than arbitrary covariance matrices.

For jump-based cases, cross-sectional dependence is induced by a common market-wide compound jump $J_t$:
\begin{equation}
\begin{aligned}
y_{i,t}
&=
\alpha_i+\phi y_{i,t-1}
+\beta_i^\top f_t+\varepsilon_{i,t}
+\gamma_iJ_t,\\
\varepsilon_{i,t}
&\sim \mathcal{N}(0,\sigma_\varepsilon^2).
\end{aligned}
\label{eq:shared_jump_panel}
\end{equation}
In the implemented Case~5 and Case~6, $f_t\equiv0$, so dependence is driven by the shared jump process and firm-level exposure $\gamma_i$.

\subsection{Case 1: Volatility Clustering with Factor and Idiosyncratic GARCH}
\label{sec:appendix_case1}

This case isolates \emph{volatility clustering}.
Both systematic factors and idiosyncratic residuals follow GARCH(1,1) conditional variance dynamics.
All nonlinear behavior arises through conditional heteroskedasticity, while cross-sectional dependence is induced by the factor structure in Eq.~\eqref{eq:shared_factor_panel}.

\paragraph{Systematic factors.}
For each factor $j=1,\dots,K$,
\begin{equation}
\begin{aligned}
f_{j,t}
&=
\mu_f+\sigma_{f,j,t}\varepsilon_{j,t},
\qquad
\varepsilon_{j,t}\sim\mathcal{N}(0,1),\\
\sigma_{f,j,t}^2
&=
\omega_f
+\alpha_f(f_{j,t-1}-\mu_f)^2
+\beta_f\sigma_{f,j,t-1}^2.
\end{aligned}
\label{eq:case1_factor_garch}
\end{equation}
We require $\alpha_f\ge0$, $\beta_f\ge0$, and $\alpha_f+\beta_f<1$.
All factors share the same unconditional variance $\bar{\sigma}_f^2$, implemented by
\begin{equation}
\omega_f
=
(1-\alpha_f-\beta_f)\bar{\sigma}_f^2.
\label{eq:case1_factor_omega}
\end{equation}

\paragraph{Idiosyncratic residuals.}
For each series $i$,
\begin{equation}
\begin{aligned}
u_{i,t}
&=
\sigma_{u,i,t}\eta_{i,t},
\qquad
\eta_{i,t}\sim\mathcal{N}(0,1),\\
\sigma_{u,i,t}^2
&=
\omega_{u,i}
+\alpha_u u_{i,t-1}^2
+\beta_u\sigma_{u,i,t-1}^2,
\end{aligned}
\label{eq:case1_idio_garch}
\end{equation}
with $\alpha_u+\beta_u<1$.
Cross-sectional heterogeneity is introduced by firm-specific unconditional idiosyncratic variances:
\begin{equation}
\begin{aligned}
\bar{\sigma}_{u,i}^2
&=
\bar{\sigma}_u^2
\exp\!\left(
\sigma_{\log}\xi_i
-\frac{1}{2}\sigma_{\log}^2
\right),
\qquad
\xi_i\sim\mathcal{N}(0,1),\\
\omega_{u,i}
&=
(1-\alpha_u-\beta_u)\bar{\sigma}_{u,i}^2.
\end{aligned}
\label{eq:case1_idio_var}
\end{equation}
The exponential normalization preserves
$\mathbb{E}[\bar{\sigma}_{u,i}^2]=\bar{\sigma}_u^2$.

\paragraph{Parameterization and simulation.}
Volatility persistence is controlled by
\begin{equation}
\rho_f=\alpha_f+\beta_f,
\qquad
\rho_u=\alpha_u+\beta_u.
\label{eq:case1_persistence}
\end{equation}
Given a fixed share $\kappa\in(0,1)$, we set
$\alpha=\kappa\rho$ and $\beta=(1-\kappa)\rho$.
We draw $(\alpha_i,\beta_i)$ once per panel, initialize conditional variances at their unconditional levels, simulate factor and idiosyncratic GARCH paths recursively, and construct returns using Eq.~\eqref{eq:shared_factor_panel}.

\subsection{Case 2: Multi-Scale Volatility Persistence via HAR Dynamics}
\label{sec:appendix_case2}

This case models \emph{multi-scale volatility persistence}, where conditional volatility depends on lagged squared shocks aggregated over multiple horizons.
The panel structure is Eq.~\eqref{eq:shared_factor_panel}; both factor and idiosyncratic components are conditionally Gaussian with HAR-style variance recursions.

\paragraph{Conditional innovations.}
For factors and residuals,
\begin{equation}
\begin{aligned}
f_{j,t}
&=
\mu_{f,j}
+\sqrt{\sigma^2_{f,j,t}}\varepsilon_{j,t},
\qquad
\varepsilon_{j,t}\sim\mathcal{N}(0,1),\\
u_{i,t}
&=
\sqrt{\sigma^2_{u,i,t}}\eta_{i,t},
\qquad
\eta_{i,t}\sim\mathcal{N}(0,1).
\end{aligned}
\label{eq:case2_innov}
\end{equation}
In implementation, $\mu_{f,j}=0$.

\paragraph{HAR-style variance recursion.}
Let $x_t$ denote a generic innovation, either a demeaned factor shock or an idiosyncratic shock.
The HAR-style variance recursion is
\begin{equation}
\begin{aligned}
\sigma_t^2
&=
c
+b_1x_{t-1}^2\\
&\quad
+b_5\frac{1}{L_{5,t}}
\sum_{\ell=1}^{L_{5,t}}x_{t-\ell}^2\\
&\quad
+b_{22}\frac{1}{L_{22,t}}
\sum_{\ell=1}^{L_{22,t}}x_{t-\ell}^2,
\end{aligned}
\label{eq:case2_har}
\end{equation}
where
\begin{equation}
L_{5,t}=\min(5,t),
\qquad
L_{22,t}=\min(22,t).
\label{eq:case2_windows}
\end{equation}
The averages include lag 1; a small floor $\epsilon>0$ ensures positivity.

For factors and residuals, the same feedback coefficients are used, but baseline levels may differ:
\begin{equation}
c_u=c_{\mathrm{idio}},
\qquad
c_f=\gamma c_{\mathrm{idio}},
\label{eq:case2_baseline}
\end{equation}
so $\gamma$ controls the factor-to-idiosyncratic baseline variance ratio.

\paragraph{Parameterization and simulation.}
Instead of specifying $(b_1,b_5,b_{22})$ directly, we use
\begin{equation}
s=b_1+b_5+b_{22},
\qquad
\lambda=\frac{b_{22}}{s},
\label{eq:case2_s_lambda}
\end{equation}
where $s\in[0,1)$ controls total feedback strength and $\lambda\in[0,1]$ controls the long-horizon share.
The implementation maps
\begin{equation}
b_{22}=\lambda s,
\qquad
b_1=b_5=\frac{(1-\lambda)s}{2}.
\label{eq:case2_coeff_map}
\end{equation}
We draw cross-sectional parameters once, initialize variances at their baseline levels, recursively update HAR variances, generate Gaussian shocks, and construct returns via Eq.~\eqref{eq:shared_factor_panel}.

\subsection{Case 3: Heavy-Tailed Innovations and Rare Outliers}
\label{sec:appendix_case3}

This case isolates distributional misspecification using factor-GARCH dynamics with heavy-tailed innovations and transient outliers.

\paragraph{Volatility dynamics.}
Factors and base idiosyncratic components follow GARCH(1,1):
\begin{equation}
\begin{aligned}
f_{j,t}
&=
\sigma_{f,j,t}z_{f,j,t},\\
u_{i,t}^{\mathrm{base}}
&=
\sigma_{u,i,t}z_{i,t},
\end{aligned}
\label{eq:case3_returns}
\end{equation}
with
\begin{equation}
\begin{aligned}
\sigma_{f,j,t}^2
&=
\omega_f
+\alpha(\sigma_{f,j,t-1}z_{f,j,t-1})^2
+\beta\sigma_{f,j,t-1}^2,\\
\sigma_{u,i,t}^2
&=
\omega_u
+\alpha(u_{i,t-1}^{\mathrm{base}})^2
+\beta\sigma_{u,i,t-1}^2.
\end{aligned}
\label{eq:case3_garch}
\end{equation}
We require $\rho=\alpha+\beta<1$ and set
\begin{equation}
\omega_f=(1-\rho)\bar{\sigma}_f^2,
\qquad
\omega_u=(1-\rho)\bar{\sigma}_u^2.
\label{eq:case3_omega}
\end{equation}

\paragraph{Heavy tails and outliers.}
To induce excess kurtosis, we use standardized Student-$t$ shocks:
\begin{equation}
z_{f,j,t},z_{i,t}
\sim
\sqrt{\frac{\nu-2}{\nu}}\, t_\nu,
\qquad
\nu>2.
\label{eq:case3_student}
\end{equation}
Rare additive idiosyncratic outliers are sampled independently:
\begin{equation}
o_{i,t}
=
\begin{cases}
s_{i,t}\cdot c_{\mathrm{out}}\sigma_{u,i,t},
& \text{with prob. }\pi_{\mathrm{out}},\\
0,
& \text{otherwise},
\end{cases}
\label{eq:case3_outlier}
\end{equation}
where $s_{i,t}\in\{-1,+1\}$ and $c_{\mathrm{out}}=\texttt{outlier\_scale}$.
The observed idiosyncratic component is
\begin{equation}
u_{i,t}
=
u_{i,t}^{\mathrm{base}}+o_{i,t}.
\label{eq:case3_obs_idio}
\end{equation}
The GARCH recursion depends only on $u_{i,t}^{\mathrm{base}}$, so the outliers represent transient distributional deviations rather than persistent volatility shocks.

\subsection{Case 4: Market-Wide Regime Switching with Structural Breaks}
\label{sec:appendix_case4}

Case~4 induces piecewise-stationary behavior through a shared latent market regime.
Let $s_t\in\{0,1,2\}$ denote the time-level regime, interpreted as Up, Stable, and Down.
Regime-specific parameters $(\mu_s,\sigma_s)$ control the conditional mean and volatility in state $s$.

\paragraph{Block-wise regime dynamics.}
To model structural breaks rather than rapid switching, the regime is constant within blocks of length $B$.
Let $T_{\mathrm{full}}=T+T_{\mathrm{burn}}$ and $b(t)=\lceil t/B\rceil$.
The block-level regimes follow a Markov chain:
\begin{equation}
\Pr(s_b=j\mid s_{b-1}=i)=\Pi_{ij},
\qquad
i,j\in\{0,1,2\},
\label{eq:case4_markov}
\end{equation}
where $\Pi$ is row-stochastic.
The time-level state is $s_t=s_{b(t)}$.

\paragraph{Regime-dependent panel process.}
Conditional on $\{s_t\}$, each series follows
\begin{equation}
\begin{aligned}
y_{i,t}
&=
a_i\mu_{s_t}
+\phi y_{i,t-1}
+b_i\sigma_{s_t}\varepsilon_{i,t},\\
\varepsilon_{i,t}
&\sim\mathcal{N}(0,1),
\end{aligned}
\label{eq:case4_panel}
\end{equation}
where $\phi\in(-1,1)$ controls within-regime persistence.
The firm-specific exposure scalars are drawn as
\begin{equation}
a_i \sim \mathrm{LN}\!\left(-\frac{1}{2}\tau_\mu^2,\tau_\mu^2\right),
\qquad
b_i \sim \mathrm{LN}\!\left(-\frac{1}{2}\tau_\sigma^2,\tau_\sigma^2\right),
\label{eq:case4_exposures}
\end{equation}
so that $\mathbb{E}[a_i]=\mathbb{E}[b_i]=1$, where $\mathrm{LN}$ denotes the log-normal distribution.

\subsection{Case 5: Market-Wide Self-Exciting Jumps}
\label{sec:appendix_case5}

Case~5 isolates self-exciting event clustering.
A market-wide event process generates clustered jump counts, and the resulting compound jump affects all series through firm-level exposures.

\paragraph{Hawkes-type intensity.}
Let $\lambda_t$ denote jump intensity and $N_t$ the event count.
The discrete-time Hawkes recursion is
\begin{equation}
\begin{aligned}
\lambda_t
&=
\mu+\delta(\lambda_{t-1}-\mu)+\alpha N_{t-1},\\
N_t\mid\lambda_t
&\sim
\mathrm{Poisson}\!\left(\max\{\lambda_t,\epsilon\}\right),
\end{aligned}
\label{eq:case5_hawkes}
\end{equation}
where $\delta=\exp(-\beta)$, $\mu\ge0$ is the baseline intensity, $\alpha\ge0$ is the excitation strength, and $\beta>0$ controls decay.
A sufficient stability condition is
\begin{equation}
\alpha<1-\delta=1-\exp(-\beta).
\label{eq:case5_stability}
\end{equation}
For reporting, we use the discrete branching proxy
\begin{equation}
\mathrm{br}_{\mathrm{disc}}
=
\frac{\alpha}{1-\exp(-\beta)},
\qquad
\mathrm{br}_{\mathrm{disc}}<1.
\label{eq:case5_branching}
\end{equation}

\paragraph{Compound jumps and panel returns.}
Conditional on $N_t$, the aggregate jump is
\begin{equation}
J_t
=
\sum_{k=1}^{N_t}s_{t,k}A_{t,k},
\label{eq:case5_jump}
\end{equation}
where $s_{t,k}\in\{-1,+1\}$ with $\Pr(s_{t,k}=+1)=p_\uparrow$.
Jump magnitudes are parameterized by their mean:
\begin{equation}
A_{t,k}\sim\mathrm{LogNormal}(m_{\log},s_{\log}^2),
\qquad
m_{\log}=\log(\bar a)-\frac{1}{2}s_{\log}^2,
\label{eq:case5_lognormal}
\end{equation}
so that $\mathbb{E}[A_{t,k}]=\bar a$.
Panel returns follow Eq.~\eqref{eq:shared_jump_panel} with $f_t\equiv0$; baseline exposures use $\gamma_i\equiv\gamma_{\mathrm{mean}}$, with optional log-normalized heterogeneity preserving the same mean.

\subsection{Case 6: Zero-Inflated Sparse Jumps}
\label{sec:appendix_case6}

Case~6 isolates sparse jump dynamics.
Unlike Case~5, jump arrivals are not self-exciting; instead, many periods are structurally inactive due to zero inflation.

Let $N_t$ denote the market-wide jump count, modeled as
$N_t\sim\mathrm{ZIP}(\pi,\lambda)$, meaning
\begin{equation}
N_t=
\begin{cases}
0,
& \text{with prob. }\pi,\\
\tilde N_t,\quad
\tilde N_t\sim\mathrm{Poisson}(\lambda),
& \text{with prob. }1-\pi.
\end{cases}
\label{eq:case6_mixture}
\end{equation}
Thus,
\begin{equation}
\Pr(N_t=0)
=
\pi+(1-\pi)e^{-\lambda},
\qquad
\mathbb{E}[N_t]
=
(1-\pi)\lambda.
\label{eq:case6_moments}
\end{equation}

\paragraph{Compound jumps and panel returns.}
Conditional on $N_t$, the common aggregate jump is again
\begin{equation}
J_t=\sum_{k=1}^{N_t}s_{t,k}A_{t,k},
\label{eq:case6_jump}
\end{equation}
with signed log-normal magnitudes as in Eq.~\eqref{eq:case5_lognormal}.
Panel returns follow Eq.~\eqref{eq:shared_jump_panel} with $f_t\equiv0$, and firm exposures are drawn once per panel:
\begin{equation}
\gamma_i
\sim
\mathcal{N}(\gamma_{\mathrm{mean}},\gamma_{\mathrm{std}}^2),
\label{eq:case6_gamma}
\end{equation}
matching the simulator's default implementation.

\end{document}